\documentclass[preprint]{aastex}

\shorttitle{Albedo Colors of Exoplanets}
\shortauthors{Cahoy et al.}
\slugcomment{ApJ accepted}

\begin{document}
\bibliographystyle{abbrevnat}


\title{Exoplanet albedo spectra and colors \\as a function of planet phase, separation, and metallicity}


\author{Kerri L. Cahoy, Mark S. Marley}
\affil{NASA Ames Research Center,
    Moffett Field, CA 94035}
\email{kerri.l.cahoy@nasa.gov}

\and

\author{Jonathan J. Fortney}
\affil{University of California Santa Cruz, Santa Cruz, CA 95064}




\begin{abstract}

First generation space-based optical coronagraphic telescopes will obtain images of cool gas and ice giant exoplanets around nearby stars.  The albedo spectra of exoplanets lying at planet-star separations larger than about $1\,\rm AU$--where an exoplanet can be resolved from its parent star--are dominated by reflected light to beyond $1\,\mu\rm m$ and are punctuated by molecular absorption features.  Here we consider how exoplanet albedo spectra and colors vary as a function of planet-star separation, metallicity, mass, and observed phase for Jupiter and Neptune analogs from 0.35 to 1 $\mu$m. We model Jupiter analogs with $1\times$ and $3\times$ the solar abundance of heavy elements, and Neptune analogs with $10\times$ and $30\times$  solar abundance of heavy elements. Our model planets orbit a solar analog parent star at separations of 0.8 AU, 2 AU, 5 AU, and 10 AU. We use a radiative-convective model to compute temperature-pressure profiles.  The giant exoplanets are found to be cloud-free at 0.8 AU, possess H$_2$O clouds at 2 AU, and have both NH$_3$ and H$_2$O clouds at 5 AU and 10 AU.   For each model planet we compute moderate resolution ($R = \lambda / \Delta \lambda \sim 800$) albedo spectra as a function of phase. We also consider low-resolution spectra and colors that are more consistent with early direct imaging capabilities. As expected, the presence and vertical structure of clouds strongly influence the albedo spectra since cloud particles not only affect optical depth but also have highly directional scattering properties. Observations at different phases also probe different volumes of atmosphere as the source-observer geometry changes. Because the images of the planets themselves will be unresolved, their phase will not necessarily be immediately obvious, and multiple observations will be needed to discriminate between the effects of planet-star separation, metallicity, and phase on the observed albedo spectra. We consider the range of these combined effects on spectra and colors. For example, we find that the spectral influence of clouds depends more on planet-star separation and hence atmospheric temperature than metallicity, and it is easier to discriminate between cloudy $1\times$ and $3\times$ Jupiters than between $10\times$ and $30\times$ Neptunes. In addition to alkalis and methane, our Jupiter models show H$_2$O absorption features near 0.94 $\mu$m.  While solar system giant planets are well separated by their broad-band colors, we find that arbitrary giant exoplanets can have a large range of possible colors and that color alone cannot be relied upon to characterize planet types.  We also predict that giant exoplanets receiving greater insolation than Jupiter will exhibit higher equator to pole temperature gradients than are found on Jupiter and thus may exhibit differing atmospheric dynamics. These results are useful for future interpretation of direct imaging exoplanet observations as well as for deriving requirements and designing filters for optical direct imaging instrumentation.

\end{abstract}






\section{Introduction} \label{sec1}

The past few years have seen the first images that resolve exoplanets from their primary stars, most notably the three planets orbiting HR 8799 seen in thermal emission \citep{mar08, ser10}.  While more such young, hot planets are sure to be seen in the coming decade, space-based coronagraphs are required to  achieve the planet-star contrast ratios ($10^{-7}$ to $10^{-10}$) needed to detect older, cooler giant planets in reflected light in the optical ($\sim 0.2$--1$\,\mu\rm m$).  Many such planets will likely be observed as part of efforts to directly image terrestrial planets at high contrast ratios, focusing on planet-star separations close enough to be within the Habitable Zone\footnote{Habitable Zone (HZ) generally refers to the luminosity-scaled equivalent of $\sim$1 AU separation from our Sun; since our focus is on giant exoplanets, we do not require a more accurate definition of the term.} of F, G, K stars like our G2V Sun.  Here we present a range of model exoplanet albedo spectra and colors with the goal of improving their interpretation and contributing to instrument design and observing plans.  We build on prior work  \citep{mar99a, sud00, sea00, for01, bur04, bur05, dyu05, sud03, sud05} by developing additional model capabilities for generating albedo spectra at different phases and by examining a wider range of plausible atmospheric compositions, planet-star separations, and planet masses and radii than previously considered. In this work, we develop a tool compatible with our existing giant planet atmospheric radiative transfer models, and do not include the effects of polarization. We refer readers interested in the use of polarimetry in the detection and characterization of exoplanets to work such as \citet{sta04, sta05, sta06, bue09}.

While numerous studies have explored the effects of system geometry and atmospheric composition on exoplanet albedo spectra for transiting close-in gas giant exoplanets \citep{for08b, rau08, spi10}, thermal emissions make significant contributions to these observations. Even the far red spectra of these close-in exoplanets can be dominated by thermal emission \citep{mar99a, bur04, for08a}. In this work, we focus instead on understanding the reflected-light spectra of widely-separated gas and ice giant exoplanets that are more likely to resemble the early direct imaging targets. 

The planets themselves will be unresolved; without additional information (e.g., radial velocity detections) the observed phase will be uncertain. This motivates us to examine the effects of planet phase as well as planet-star separation, mass, and metallicity on the reflected-light albedo spectra of Jupiter and Neptune analogs. We use the term `metallicity' to refer to an enhancement in heavy elements compared to solar elemental abundances.

\subsection{Contrast and phase} \label{subs1_1}

The contrast ratio between an exoplanet and its star depends on many factors, including the physical properties of the planet and star, the system geometry, background sources, and the instrumentation used. A simple way of estimating the required contrast is to assume that the planet reflects its incident starlight isotropically, such that its apparent brightness is constant over its illuminated surface, as a Lambertian surface. Then the contrast can be expressed as \citep{sob75}:

\begin{equation}
C\left(\alpha\right) = \frac{2}{3}\rm A_g\left(\lambda\right)\left(\frac{R_p}{\it d}\right)^2\left[\frac{\sin\alpha+\left(\pi-\alpha\right)\cos\alpha}{\pi}\right]
\label{eq_contrast}
\end{equation}

\noindent where $\alpha$ is the planet phase angle, $R_p$ is the planet radius in km, $d$ is the planet-star separation in km, and $\rm A_g$ is the geometric albedo, which we formally define in \S \ref{subs2_2}. The geometric albedo generally takes values between 0 and 1 for the fraction of monochromatic light the planet reflects towards the observer at full phase, although it can be larger than 1 for anisotropic scattering atmospheres or surfaces. If one assumes the planet is at quadrature, $\alpha=\pi/2$, and that the geometric albedo at the wavelength of interest is $0.5$, then for a Jupiter analog ($1\, \rm R_J$), the contrast at 0.8 AU is $\sim3.8\times10^{-8}$ and at 10 AU is $\sim2.4\times10^{-10}$. For a Neptune analog ($1\, \rm R_N$), the contrast at 0.8 AU is $\sim4.5\times10^{-9}$ and at 10 AU is $\sim3\times10^{-11}$. For an Earth analog ($1\, \rm R_E$) at 1 AU with $\rm A_g = 0.3$, the contrast is $\sim1.2\times10^{-10}$. High contrast is more difficult to achieve closer to the star than further away, however, the $d^{-2}$ dependence also means that more distant objects will be fainter in reflected light.

\subsection{Composition} \label{subs1_2}

As discussed in \citet{mar07}, determining atmospheric composition is a central objective for characterizing extrasolar giant planets.  A census  of how composition varies with planet mass and orbital distance will help inform theories of giant planet formation. The giant planets in our solar system motivate us to consider a large range of metallicities in our model exoplanets. As summarized in \citet{for08a}, the \textit{Galileo} and \textit{Cassini} spacecraft have shown that the atmospheres of Jupiter and Saturn have enhanced heavy element abundances compared to that of the Sun. For example, Jupiter has $\sim 2\times$ to $4\times$ solar abundances of oxygen, carbon, nitrogen, sulfur, and various noble gases \citep{atr03}. Understanding how giant planets formed and the role of processes such as planetesimal bombardment and accumulation during formation \citep{owe99, gau01a, gau01b, gui00, ali05}, erosion of the heavy-element core \citep{ste85, gui04}, direct accretion of metal-rich disk gas \citep{gui06}, or chemical fractionation within the planet \citep{ste77, lod04} in enriching gas giants in heavy elements is relevant both to understanding our own solar system as well as exoplanet systems.

This motivates our goal of better understanding the combined effect of planet-star separation, composition and planet phase on spectra that are likely to be obtainable in early reflected-light images of exoplanets. For this initial study, we assume simple circular orbits where the planet-star separation is constant, and examine how the albedo spectra and colors change with planet phase due to the change in angle from the source to the observer. More complex orbits, such as eccentric orbits, require consideration of how the structure of the exoplanet's atmosphere would evolve as a function of distance from its parent star.  An initial study of the effect of eccentric orbits, using interpolated temperature and pressure profiles corresponding to different planet-star separations, was presented for a  $1\,\rm M_J$ extrasolar giant planet (EGP) in \citet{sud05}.  

\subsection{Clouds} \label{subs1_3}

As with any planet, exoplanet albedo spectra are very sensitive to the presence and structure of clouds  \citep{mar99a, sud00}. Because the global height and optical thickness of cloud layers are difficult to model on an a priori basis, \citet{mar99a} considered a few characteristic cloud models in order to test the sensitivity of albedos to clouds.  \citet{sud00} likewise employed a fairly simple method for computing cloud opacity.  In this study, we use the more complex cloud model of \citet{ack01} to calculate the height, particle sizes and consequently optical thickness of water and ammonia clouds.  A comparison of the \citet{ack01} model with the models used in \citet{mar99a} can also be found in \citet{ack01}. The \citet{ack01} cloud model provides wavelength-dependent measures of the anisotropic scattering in terms of the asymmetry factor $g$, and single-scattering albedo $\tilde{\omega}$ of the cloud particles at each temperature and pressure level. In \S \ref{sec3} we discuss how these parameters are used to approximate anisotropic scattering as well as describe planned modifications to the cloud model to provide additional scattering information. 

The \citet{sud00, sud05}  models generally focused on $1\,\rm M_J$ planets and considered the effects that might be manifested in planets with inclined or highly elliptical orbits (such as transitioning from cloudy to cloud-free and thus bright to dark in reflected light). The effect of geometry on the orbit determination of close-in exoplanets from photometry for was also recently presented in detail by \citet{bro09}. Here we extend the work of \citet{mar99a} to model exoplanet observations as a function of phase and \citet{sud00, sud05} by considering a greater range of planet masses and metallicities.  In this work we consider both cloud-free (0.8 AU) and cloudy ($> 1$ AU) planets. 

\subsection{Organization} \label{subs1_4}

We briefly review the basics of exoplanet albedo theory in \S \ref{sec2} as framework for the following results and discussion. While this material is familiar to the solar system planetary science community, it bears reinforcing because of the subtleties between various types of planetary albedos and their relationship to phase.  In \S \ref{sec3} we describe how we generate a series of exoplanet model atmospheres over a range of planet-star separations and metallicities using a 1-D radiative-convective model that was previously tailored for use with EGPs \citep{mar97, for05, for06, mar07, for07b, for08a, for08b} and brown dwarfs \citep{mar96, bur97, mar97, mar02, sau06, sau07}. The model was originally used for solar system planetary bodies such as Titan and Uranus \citep{too77, too89, mck89, mar99b}. These global mean 1D exoplanet models are then used as input to a high resolution albedo spectra model. We describe our updates to the albedo model that allow us to calculate emergent intensities and thus albedo spectra as a function of phase.  

In \S \ref{sec4} we present results: albedo spectra from 0.35 $\mu$m to 1 $\mu$m as a function of planet type, planet-star separation, metallicity, and phase. Because of the practical constraints inherent in spacecraft flybys of solar system giant planets, there is relatively little data in the literature reporting planet-averaged phase functions; we do compare Voyager 1 data of Uranus and Neptune from \citet{pol86} with our model phase functions. We also compare the albedo spectra with observations of solar system giant planets from \citet{kar94}. Since early exoplanet direct imaging observations will have limited resolution, in \S \ref{sec5} we consider lower-resolution, coarse spectra derived from our high-resolution results. We consider $R =  \lambda /\Delta \lambda = 5$ and 15, and we also consider how changes in albedo spectra might present themselves in terms of color-color comparisons using standard filters in the optical. This helps us to evaluate features that might be detectable using a combination of different wide (or narrow) filter bands. 

Color-color diagrams were suggested as a comparative analytic tool for direct imaging of exoplanets in an example that considered the color diversity of solar system planets in \citet{tra03}. Recently, \citet{for08a} presented detailed color-color comparisons of hot young Jupiters in the infrared. Earlier, \citet{sud05} modeled reflected light albedo spectra and performed color-color analyses in the optical of a $1\, \rm M_J$ EGP using a similar approach but with a different model and implementation than that used in this work. While our $1\, \rm M_J$ results that include clouds tend to be less red at full phase and less blue at new compared with those in \citet{sud05}, we generally agree with their cloud-free result and extend their approach to include different compositions, Neptune analogs, and comparisons with observed spectra and colors of solar system planets.


\section{Background} \label{sec2}

This section provides some background on exoplanet direct imaging methods and the scientific value of albedo spectra constructed from observations made in the optical. We briefly review common direct imaging terminology and techniques in \S \ref{subs2_1} and then formally define geometric albedo and illustrate how albedo spectra of exoplanets relate to exoplanet science goals in \S \ref{subs2_2}. We also briefly discuss the challenges and current limitations that affect direct imaging efforts. Consideration of both the instrumentation constraints for direct imaging as well as exoplanet science goals help to define and justify the planet-star separations and atmospheric model types used in this work.


\subsection{Direct Imaging} \label{subs2_1}

\begin{figure}
\epsscale{1}
\plotone{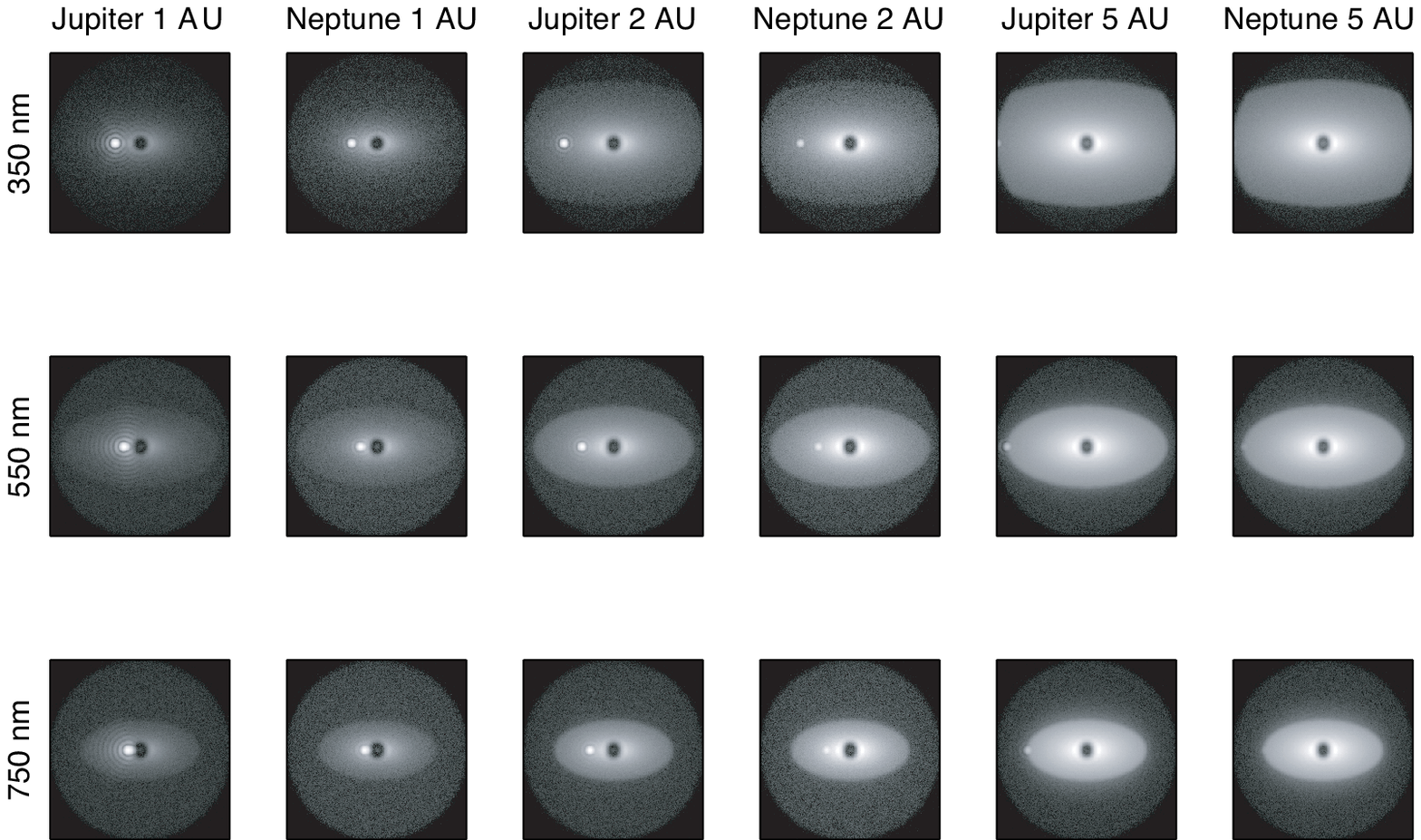}
\caption{Simulated direct images of Jupiters and Neptunes around a solar analog at 10 pc, as a function of planet-star separation (1 AU, 2 AU, and 5 AU) and $\lambda$ (350 nm, 550 nm, and 750 nm, each with 100 nm bandwidth). The simulation is for a D = 4 m space telescope with a Phase-Induced Amplitude Apodization (PIAA) coronagraph \citep{guy05}. The PIAA coronagraph has an inner working angle $\sim2\lambda/D$. The integration time is 10 hours and effects from photon noise and 1 zodi of both local and exozodiacal dust are included (system inclination is $60^{\circ}$). Note the scale with $\lambda$ in these 256$\times$ 256 pixel images: 350 nm $\sim$2.8 mas/pix, 550 nm $\sim$4.4 mas/pix, and 750 nm $\sim$6.0 mas/pix \citep{cah09}. \label{coropics}}
\end{figure}

To give the reader an idea of the challenges inherent in coronagraphic detection and characterization of gas or ice giant exoplanets, Figure \ref{coropics} shows simulated direct images of Jupiter and Neptune analogs observed in three different 100 nm wide bands around a solar analog at a distance of 10 pc from the observer \citep{cah09}. The simulations are performed for a 4 m diameter space telescope, where the light from the parent star has been suppressed by a Phase Induced Amplitude Apodization (PIAA) coronagraph \citep{guy05}, and stellar leakage is included as one of several background sources. Importantly, for this figure only, we assume a \textit{gray} albedo spectrum with a geometric albedo that is a constant 0.3 with wavelength and thus the same for each planet in each bandpass. The simulations are described in greater detail in \citet{cah09}.

Figure \ref{coropics} helps to emphasize the importance of planet-star contrast ratios and angular resolution in a direct imaging observation. It also puts into context several factors that are not specific to a particular instrument or method, but that are important to consider when both interpreting observations and when constructing exoplanet models that correspond to early direct imaging capabilities. For example, the simulated images make clear the relevance of planet size, wavelength-dependent brightness (albedo spectra, one can imagine the effect of a decreasing albedo with wavelength on the images), minimum and maximum detectable separations from the parent star, contributions from background sources like exozodiacal debris disks, and the effect of system geometry and planet phase on the observations. The wavelength dependent plate scales in this example also give a sense of the level of calibration required to construct coarse albedo spectra even from concurrent measurements. For photon-starved exoplanet observations, the tradeoff between resolution, integration times, and noise is the major challenge in obtaining data in multiple bands. 

The term inner working angle (IWA) is often used to describe the minimum detectable separation between a star and planet in a high-contrast direct imaging measurement, where the IWA is generally substantially larger than the diffraction limit (e.g., the 1.22$\lambda/D$ Rayleigh criterion). For a band-limited coronagraph, the IWA is defined as the separation at which the planet's light is at half of the maximal throughput of the starlight suppression technique at a contrast level sufficient for detection \citep{cag09}. An instrument that is improved to achieve a smaller IWA can detect exoplanets at closer star-planet separations and thus can also detect exoplanets around stars at greater distances from the observer. The IWA for current observing systems is typically on the order of a few tenths of an arcsecond \citep{bei10}; instruments in development may achieve IWAs on the order of a tenth of an arcsecond. Examples of some current direct imaging methods include high-performance coronagraphs \citep{kuc02, kas03, guy05, traug07, maw09}, external occulters \citep{cas06}, and optical nullers \citep{wal00, mle06}, as well as post-processing algorithms such as those used by \citet{mar06} and \citet{laf07}.

In practice, for current direct imaging methods the IWA is typically larger than $2\lambda/D$ at the required contrast levels for imaging exoplanets in reflected light (the IWA for the PIAA coronagraph simulated in Fig.~\ref{coropics} is also $\sim2\lambda/D$). Early optical direct images of exoplanets will likely be limited to planet-star separations greater than $2\lambda/D$, since the IWA metric generally does not include external contributing factors such as noise, background sources, coatings, optical throughput, quantum efficiency, etc. There are also platform-specific limitations, for example, primary mirror diameter $D$ is limited on space-based direct imaging missions by either deployment complexity, cost, or launch vehicle fairing size. Ground-based instruments do not have this limitation on $D$, however, they do have other challenges, such as contrast limitations due to turbulence in Earth's atmosphere. Given the capability of current technology, using a minimum separation of $\sim1$ AU for our exoplanet models around a Sun-like star seems reasonable for this first set of simulations. In addition, a separation of $\sim1$ AU is far enough away from a Sun-like star that reflected light will dominate the features in the optical spectrum, even of gas giants \citep{mar99a}.

The model planets shown in Fig.~\ref{coropics} are based on their solar system counterparts and are separated from their parent star by luminosity-scaled equivalents of 1 AU, 2 AU, and 5 AU. We hold the albedos constant vs. $\lambda$ at 0.3 in these simulations, to better illustrate the other wavelength-dependent changes. If we were to ignore the wavelength-dependent changes in albedo, the larger-radius gas giant planets are simply brighter and easier to detect than terrestrial exoplanets. This is particularly true at redder wavelengths where the $\lambda$ dependence of the IWA `pushes' the planet closer to the IWA, effectively making smaller separations harder to detect at longer wavelengths. As we discuss in \S \ref{sec4}, realistic giant planets get darker at smaller planet-star separations when the temperatures become too warm for bright clouds to form. Depending on the star, the darkening at temperatures that are too high for cloud formation may occur inside of the IWA, making the detection of such planets less likely. For reflected-light direct imaging of cooler exoplanets, Fig.~\ref{coropics} show that it is easier to detect the larger Jupiter and Neptune analogs compared with terrestrial planets, which motivates our use of them as the exoplanet models in this work.

For some direct imaging methods, there may also be limitations on the maximum planet-star separation, or outer working angle (OWA). This can be due to the optical design or elements such as deformable mirrors, detector size, sub-apertures, etc. It is desirable to have an OWA of $\sim20\lambda/D$ or larger, if possible, so that in addition to exoplanets, exozodical dust and circumstellar debris disk structure can be characterized at much wider separations.

The existence of giant planets at separations larger than $\sim10$ AU is difficult to account for in standard core accretion models \citep{pol96, ida05, dod09}. Their presence at very large planet-star separations may be due to a different formation mechanism, such as gravitational fragmentation in the disk \citep{bos00, kra10} or due to a combination of multiple effects that move planets formed in dense inner regions of a disk out to distant orbits, including outward migration and planet-planet scattering \citep{ver09}. 

In this work,  we model Jupiters and Neptunes out to 10 AU based on the theoretical predictions for giant planet separations and the fact that the contrast requirement for cool giant planets beyond 10 AU is challenging ($d^{-2}$ dependence on the incident light, as noted in \S \ref{subs1_1}).  Using Eq.~\ref{eq_contrast}, a $1\, \rm R_J$ planet with geometric albedo of 0.5 has a contrast of $\sim1\times10^{-10}$ at 15 AU and $\sim6\times10^{-11}$ at 20 AU when observed in reflected light at quadrature. Such contrasts are comparable to or fainter than that expected for terrestrial planets at 1 AU, thus motivating our outer model boundary at 10 AU.  In \S \ref{sec3} we discuss the role that planet-star separation plays in controlling the atmospheric structure (clouds) and albedo spectra of our model exoplanets.

\begin{figure}
\epsscale{1}
\plottwo{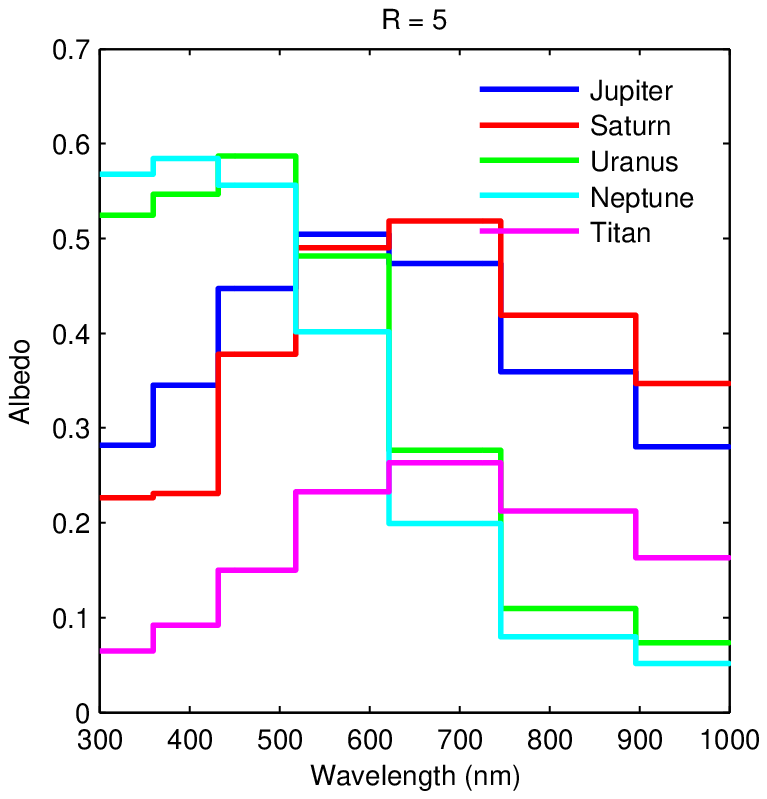}{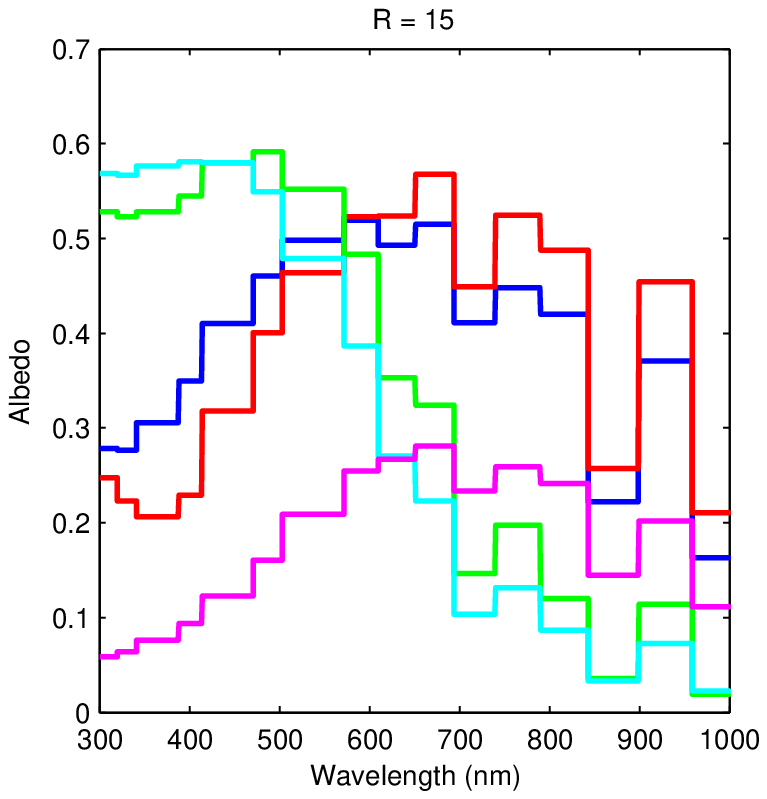}
\caption{Low-resolution albedo spectra of the outer planets in our solar system. Original high-resolution data from \citet{kar94}. Left, $R = \lambda/\Delta \lambda = 5$, right, $R = 15$. \label{lowreskar}}
\end{figure}

Simultaneous images taken with different filters can function as coarse spectra. In Fig.~\ref{lowreskar} we use observed spectra of the outer solar system planets and Titan from \citet{kar94} to illustrate how coarse spectra from an early direct imaging observation might be used to differentiate between planet types. The original spectral data from \citet{kar94} are binned down to resolution $R = \lambda/\Delta \lambda = 5$ and $R = 15$. Even in the $R = 5$ spectra,  one can distinguish between the ice giants (Neptune, Uranus) and the gas giants (Jupiter, Saturn). The ice giants are brightest in the blue and get darker in the red, and the gas giants are not as bright in the blue, due partly to the effects of photochemical products, but are brighter mid-band and in the red.  An important goal of our study is to understand the extent to which such distinctions hold in general for exoplanets and how the phase at which they are observed affects the colors.

As we discuss in more detail in \S \ref{subs2_2} and \S \ref{sec4}, coarse spectra can be used in combination with atmospheric models to constrain the temperature, composition, the presence or absence of clouds in an exoplanet's atmosphere, and the exoplanet's radius \citep{mar99a, mar99b, mar07, for08a, for08b}. Color-color comparisons of exoplanets can be made \citep{tra03, sud05, for08a}, and understanding how system geometry and explanet composition relate to variability in the observations is important for interpreting these color-color comparisons.


\subsection{Albedos} \label{subs2_2}

The definition and relationships between between geometric albedo $\rm A_g$, phase $\alpha$, the phase integral $q$, spherical albedo $\rm A_s$, and Bond albedo $\rm A_B$, and additional parameters of interest such as planetary effective temperature are important to understand in the context of interpreting direct imaging observations.  Although this material is not new, we review it since the distinctions between the various types of albedos are important to emphasize.  Note that while \citet{mar99a} presented geometric albedo spectra and tables of Bond albedos for exoplanets in systems with varying stellar types, our focus here is more closely on geometric albedo, albedo spectra, and planet phase functions. We do tabulate Bond albedos for our exoplanet models, but only for a solar analog parent star. 

Measurements of the planet's brightness at a given wavelength and \emph{full phase} yields the planet's monochromatic geometric albedo:

\begin{equation}
\rm A_g\left(\lambda\right)  = \frac{F_{p}\left(\lambda, \alpha=0^{\circ}\right)}{F_{\odot,L}\left(\lambda\right)}
\label{eq1}
\end{equation}

\noindent Geometric albedo $\rm A_g\left(\lambda\right)$ is the ratio of the reflected flux $\rm F_{p}\left(\lambda, \alpha=0^{\circ}\right)$ of an object at full phase ($\alpha = 0^{\circ}$) to the flux from a perfect Lambert disk, $F_{\odot,L}$of the same radius $R_p$ under the same incident flux $\rm F_{\odot}$ at the same distance $d$ from the star. A planet's phase angle $\alpha$ is the angle between the incident ray from the star to the planet and the reflected ray from the planet to the observer. The scattering angle $\Theta$ is related to the phase angle via $\alpha = \pi - \Theta$. 

In this work, we use the term \emph{geometric albedo} to refer to albedo spectra at full phase. We use the term \emph{albedo spectra} to refer to spectra observed at different orbital phases.

In order to reduce confusion in later discussions of planet phase functions and particle scattering functions, we use the notation $\rm A_g\left(\lambda\right)$ for geometric albedo here instead of using the other common notation for geometric albedo, $p$. As noted in \citet{sob75}, for a perfectly reflecting Lambert sphere, the geometric albedo is 2/3, and for a semi-infinite purely Rayleigh scattering atmosphere, it is 3/4 \citep{dlu74}. 

Flux measurements at different planet phase angles can be used to define the phase function $\Phi\left(\lambda,\alpha\right)$:

\begin{equation}
\frac{\rm F_p\left(\lambda,\alpha\right)}{\rm F_\odot\left(\lambda\right)} = \rm A_{g}\left(\lambda\right)\left(\frac{\it R_p}{\it d}\right)^2 \Phi(\lambda,\alpha)
\label{eq2}
\end{equation}

\noindent As noted for Eq.~\ref{eq1}, $\rm A_{g}\left(\lambda\right)$ is defined at $\alpha = 0^{\circ}$, $\rm F_p\left(\lambda,\alpha\right)$ is the monochromatic planet flux, $\rm F_\odot\left(\lambda\right)$ the monochromatic stellar flux, $R_p$ the planet's radius, $d$ the planet-star separation, and $\Phi(\lambda,\alpha)$ the planet's phase function at angle $\alpha$. The phase function is normalized to be 1.0 at full phase. The scattering phase function of the particles in the exoplanet's atmosphere, $p\left(\Theta\right)$ contribute to the disk-integrated phase function. The observed phase function allows calculation of the phase integral, $q$:

\begin{equation}
q\left(\lambda\right) = 2 \int_0^{\pi} \Phi(\lambda,\alpha)\sin\alpha d\alpha
\label{eq3}
\end{equation}

\noindent The spherical albedo,

\begin{equation}
\rm A_s\left(\lambda\right) =\it q\left(\lambda\right)\rm A_g\left(\lambda\right)
\label{eq4}
\end{equation}

\noindent is the fraction of incident light reflected towards all angles.

The planet's Bond albedo, an incident flux-weighted and wavelength-integrated function of $\rm A_s\left(\lambda\right)$ and $\Phi(\lambda,\alpha)$, normalized by the incident stellar flux, can then be computed:

\begin{equation}
\rm A_B = \frac{\int_0^{\infty}\rm A_{s}\left(\lambda\right)\rm F_{\odot}\left(\lambda\right)d\lambda}{\int_0^{\infty}\rm F_{\odot}\left(\lambda\right) d\lambda} 
\label{eq5}
\end{equation}

\noindent  It is important to appreciate that since the Bond albedo is weighted by the incident flux, two identical planets with identical geometric albedo spectra will have different Bond albedos if the incident flux is different (e.g., comparing an F star and an M star) \citep{mar99a}.  The Bond albedo can be used to find the effective temperature, $\rm T_{eff}$ of a rapidly rotating planet:

\begin{equation}
4\pi R_p^2 \sigma \rm T_{eff}^4 = E_{int} + \pi \it R_p^2\left[f\left(\rm1- A_B\right)\right]\frac{\pi \it S}{d^2} 
\label{eq6}
\end{equation}

\noindent where $\sigma$ is the Steffan-Boltzmann constant, $\pi S$ is the incident flux of the parent star at a planet-star separation of 1 AU, $d$ is the planet-star separation, here in AU, and $E_{int}$ represents the internal energy sources of the planet \citep[e.g.][]{con89}.  

An additional redistribution parameter $f$ may also be introduced to parameterize the global redistribution of energy to solve for a day and nightside temperature. For solar system giant planets, typically $f = 1$, since Jupiter or Neptune-like planets with rotation periods of hours and effective temperatures below 500 K have sufficiently long radiative time constants such that atmospheric dynamics likely to redistribute incident flux around the planet. For exoplanets, sometimes different values of $f$ are used due to the diversity of exoplanet system geometries. For example, $f = 2$ represents the case where the redistribution of the absorbed flux is over the day side only, yielding a higher day side effective temperature.  Here, we use $f = 1$.

In \S \ref{sec3} and Appendix \ref{App} we describe updates to our albedo spectra model that allow us to generate albedo spectra at different planet phases for a range of input model gas and ice giant exoplanet atmospheres, which can then be used to calculate exoplanet colors, phase integrals, spherical albedos, and Bond albedos.




\section{Description of Methods and Models} \label{sec3}

In this section, we describe the two-part approach to generating our results.  For each model planet we first compute a mean thermal structure that accounts for the deposition of incident flux and an assumed internal heatflow.  The 1D mean model is computed utilizing our radiative-convective equilibrium model accounting for transport of thermal emitted flux and incident light.  Once we have the mean global structure in hand we compute the detailed scattering calculations as a function of orbital phase using our albedo model.

In this section, we describe the two-part approach to generating our results. For each model planet we Þrst compute a mean thermal structure that accounts for the deposition of incident ßux and an assumed internal heatßow. The 1D mean model is computed utilizing
our radiative-convective equilibrium model accounting for transport of thermal emitted ßux and incident light.  Because radiative timescales are longer than a rotation period and atmospheric and internal dynamics tends to distribute energy around the planet \citep{ing78}, the 1D mean profile is expected to be an excellent description of the atmospheric profile at most points on the planet \footnote{Note the discussion in \S \ref{subs4_3a} speculating that some giant planets may have cold polar regions; such a temperature pattern would break our assumption of a homogeneous global thermal profile.}. Once we have the mean global structure in hand we compute the detailed scattering calculations as a function of orbital phase using our albedo model. The 1D model alone could not be used for this latter application since it treats mean conditions encountered
at a Þxed stellar zenith angle. Our two step approach allows detailed modeling of the scattered light at arbitrary phase and spectral resolution and follows our similar approaches at Titan and Uranus with earlier versions of the same models \citep{mck89, mar99b}.  For clarity we will use the description `radiative-convective model' to refer to the model employed to generate the 1D exoplanet models and the description `albedo model' to refer to the model used to generate the albedo spectra. 


\subsection{Radiative-Convective Exoplanet Model} \label{subs3_1}

We generate the exoplanet model atmospheres with the 1-D iterative radiative-convective  atmosphere model previously used in \citet{mar97,bur97, mar99a, mar02, for05, for06, for07a, for07b} and \citet{for08a, for08b}. The radiative transfer method was developed by \citet{too77, too89}, implemented by \citet{mck89} and \citet{mar99b} and used in application to Titan \citep{mck89} and Uranus \citep{mar99b} as well as to EGPs and brown dwarfs, as noted in \S \ref{sec1}.  The model computes the deposition of incident flux from $0.268\,\rm \mu m$ to $6.064\,\rm \mu m$  and accounts for thermal emission out to $227\,\rm \mu m$. 

In this work, we use elemental abundances from \citet{lod03}.  We employ tables of chemical equilibrium compositions following \citet{feg94} and \citet{lod02, lod06, fre08}. The chemistry calculations include `rainout,' where refractory species are depleted from the atmosphere due to their condensation into cloud decks \citep{lod99}. The spectra of both brown dwarfs and our solar system's giant planets can only be reproduced when chemistry calculations incorporate this process \citep{feg94, mar02, bur02}. 

As noted in \S \ref{subs1_1}, the four gas giants in our solar system show an enhancement of heavy elements compared to solar abundances. The amount of enhancement provides additional clues about planet formation and evolution. Our models include a range of compositions, with metallicities of $1\times$ and $3\times$ solar for the model Jupiters, and $10\times$ and $30\times$ solar for the model Neptunes. To be specific, we use metallicities of $\rm \left[M/H\right] / [M/H_{\rm cosmic}]= 0.0$, 0.5, 1.0, and 1.5.  So although we refer to them as $3\times$ and $30\times$, the $3\times$ and $30\times$ metallicities are actually $3.16\times$ and $31.6\times$. 

The radiative-convective model uses the correlated-\emph{k} method for the tabulation of gaseous opacities \citep{goo89}. The extensive opacity database we use is fully described in \citet{fre08}. We note that we have yet to incorporate the recent tabulation of methane opacities presented in \citet{kar10}. We also note the general difficulties in determining the accuracy of opacities at the temperature and pressures of interest, and of reconciling laboratory observations with Earth-based, space-based, and in-situ observations. 

We use the cloud model of \citet{ack01} to describe the location, vertical distribution, and particle sizes of the major cloud-forming species, which are 
H$_2$O and NH$_3$ in this work.  Clouds are coupled iteratively into the radiative-convective model using the approach of \citet{ack01} as described in \S \ref{subs1_3}.  The model computes the vertical variation of particle sizes and number densities given an assumed cloud sedimentation efficiency, $f_{\rm sed}$ which we discuss further below.  Since the  cloud model is  converged along with the atmospheric thermal structure, the clouds are fully self consistent with the atmospheric thermal structure of each model planet.  Note that we neglect $\rm NH_4SH$, which may produce an intermediate cloud between the water and ammonia clouds in Jupiter's atmosphere \citep{wes04}.  Even at Jupiter the relative roles that ${\rm NH_3}$ and ${\rm NH_4SH}$ clouds play in influencing the albedo spectra remain uncertain \citep{irw05} and thus we treat only the former cloud.

For each model planet our 1D modeling thus produces a vertical temperature and pressure, T(P), profile, a cloud structure model, and the variation in composition of all major species.
We use these profiles and cloud information (wavelength-dependent asymmetry factor $g$ and single-scattering albedo $\tilde{\omega}$ at each temperature and pressure level) as input to the  albedo spectral model described in \S \ref{subs3_2}.  

\subsubsection{Model Temperature and Pressure Profiles} \label{subs3_1a}

Table \ref{tab1} summarizes the exoplanet model gas giants (Jupiters) and ice giants (Neptunes) used in this work.  The internal heat flows assumed for the Jupiters and Neptunes correspond to effective temperatures ($T_{\rm{int}}$) of 100 K and 50 K, respectively, for an isolated object, and an age of $\sim$4.5 Gyr. Gravity for Jupiters is 25 ms$^{-2}$ and for Neptunes is 10 ms$^{-2}$. We compute radiative-convective equilibrium T(P) profiles and low-resolution spectra for exoplanet models at planet-star separations of 0.8 AU, 2 AU, 5 AU, and 10 AU at $1\times$ and $3\times$ solar metallicity for the Jupiters and $10\times$ and $30\times$ solar metallicity for the Neptunes. As noted in \S \ref{subs2_1}, the maximum and minimum planet-star separations were chosen to span the range of exoplanets detectable with early coronagraphic direct imaging methods. 

\begin{figure}
\epsscale{1}
 \plottwo{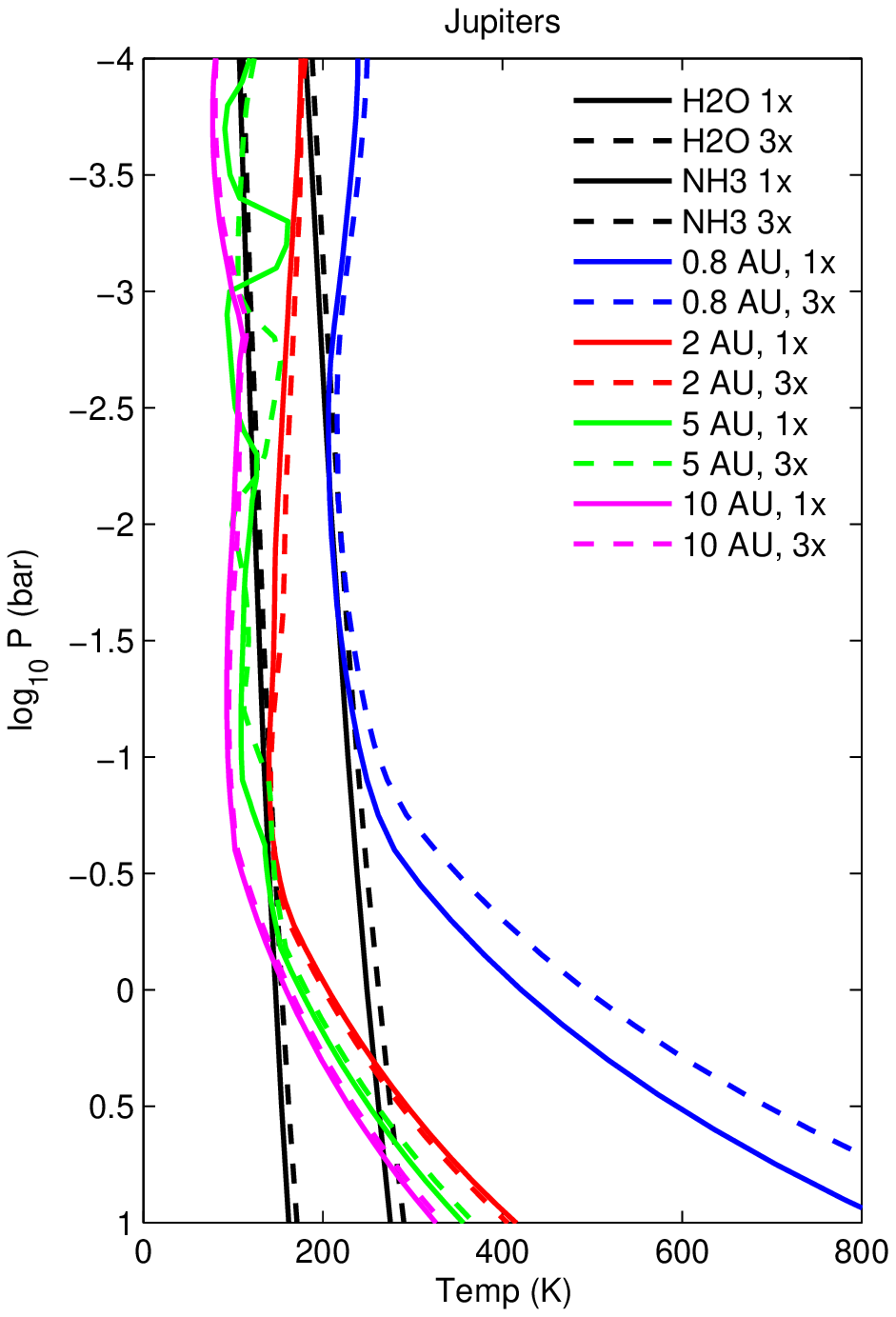}{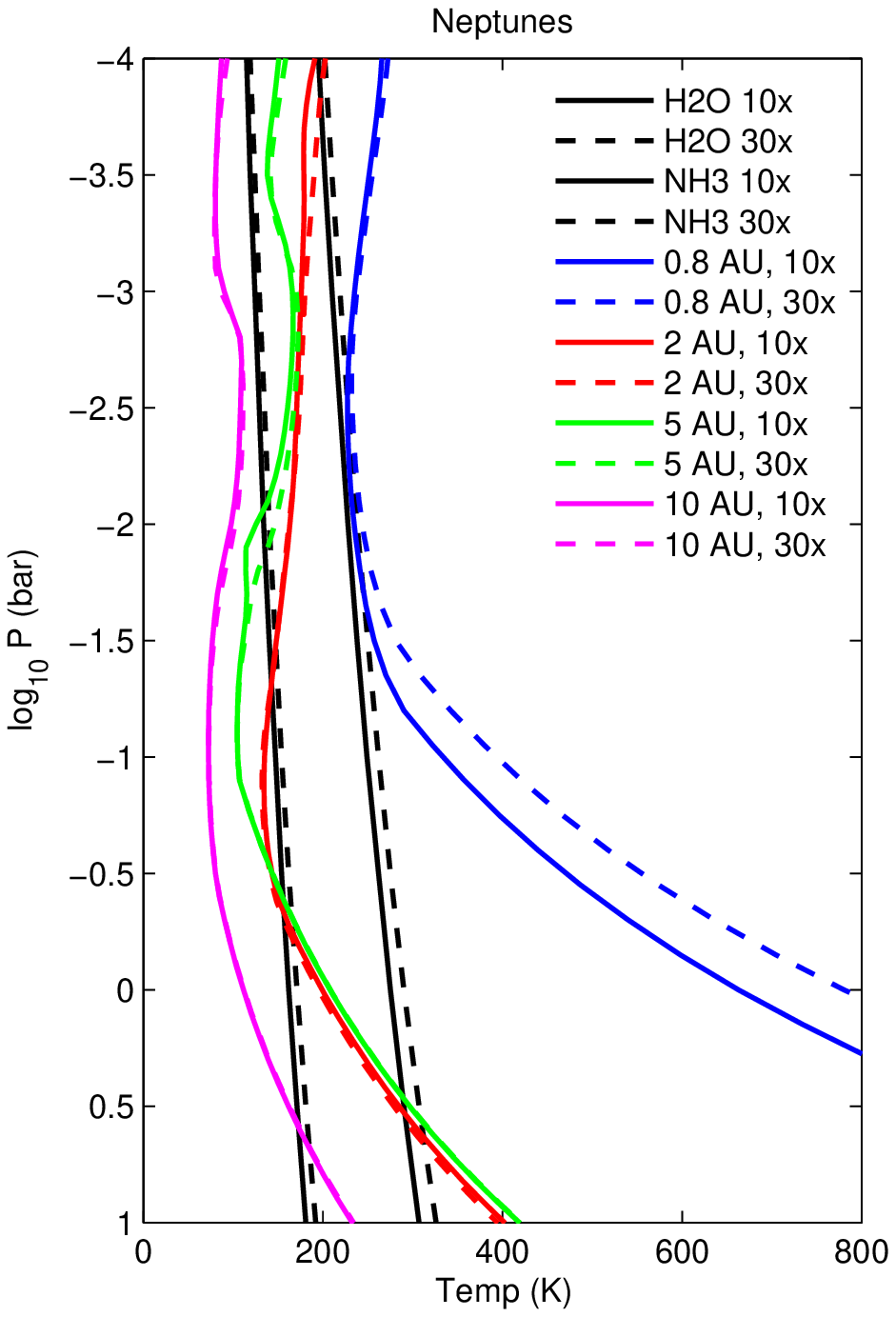}
\caption{Temperature and pressure models used as input to the albedo spectra model. Left, models for Jupiter with compositions of 1$\times$ and 3$\times$ solar and at separations of 0.8 AU, 2 AU, 5 AU, and 10 AU. Right, models for Neptune with compositions of 10$\times$ and 30$\times$ solar and at separations of 0.8 AU, 2 AU, 5 AU, and 10 AU. See additional model descriptions in \S \ref{sec3} and Table \ref{tab1}. \label{temppres}}
\end{figure}

Figure \ref{temppres} shows the resulting T(P) profiles for the models. Solid lines are used for the lower metallicity $1\times$ and $10\times$ cases, and dashed lines are used for the higher metallicity $3\times$ and $30\times$  cases. Condensation curves, courtesy K.\ Lodders, are shown on each plot for both H$_2$O (warmer) and NH$_3$ (cooler). The condensation curves correspond to $1\times$ (solid line) and $3\times$ (dashed line) solar metallicity for the Jupiter models and $10\times$ (solid line) and $30\times$ (dashed line) solar metallicity for the Neptune models.

To illustrate the impact of clouds on the albedo spectra and thus the detectability of exoplanets, we chose planet models across the full range of planet-star separations likely to be probed by early direct imaging observations. Our planet-star separations include cloud-free, H$_2$O, and combined H$_2$O and NH$_3$ cloud cases for both the Jupiters and Neptunes. However, for the purposes of comparison and simplicity, we also used the same planet-star separations for both the Jupiters and the Neptunes to limit the number of models to a few representative cases. 

While the \citet{ack01} approach results in clouds with varying particle sizes as a function of height through the atmosphere, for heuristic purposes the mean or `effective' cloud particle radii can also be computed (see \citet{ack01}).  Table \ref{tab1b} lists the effective radii of the condensate particles for the clouds in each of the exoplanet models.  We stress that these are provided here for comparison between the models; the actual radiative transfer calculations do not employ these mean particle sizes, but rather the actual computed particle size profiles.  We used the same values for $f_{sed}$, the ratio of the microphysical sedimentation flux to the eddy sedimentation flux, for both Jupiter and Neptune models at a given planet-star separation. At 2 AU, we used $f_{\rm sed} = 6$, and at 5 AU and 10 AU, we used $f_{\rm sed} = 10$.  These values of $f_{\rm sed}$ were selected partly from our experience with fitting L dwarf spectra \citep{ste09} and partly to allow convergence (very thick clouds modeled with low values of $f_{\rm sed}$ do not converge well).  \citet{ack01} found good agreement with the NH$_3$ cloud in Jupiter's atmosphere using $f_{\rm sed} = 3$.  Larger values of $f_{\rm sed}$ lead to clouds of slightly smaller vertical extent and optical depth \citep{ack01}. In a future study, we will explore the effect of changing cloud thickness on planet colors.  For the current study, all of the clouds have large optical depths, so the impact of varying $f_{\rm sed}$ is small. We stress that for this study we are primarily interested in the macro effect of clouds and not in detailed changes to the cloud model. A higher value of $f_{\rm sed} $ was required for the water cloud planets in order to produce properly converged models.

Fig.~\ref{temppres} shows that at  0.8 AU, both the $1\times$ and $3\times$Jupiters are just warm enough to be cloud-free. At 2 AU the Jupiters are just warm enough to only condense H$_2$O clouds, and at 5 AU and 10 AU the Jupiters have both H$_2$O and NH$_3$ clouds. Fig.~\ref{temppres} also shows that the H$_2$O and NH$_3$ condensation curves for the $10\times$ and $30\times$ Neptunes are a bit warmer than their $1\times$ and $3\times$ counterparts. The 0.8 AU Neptunes are also just warm enough to be cloud-free. The 2 AU Neptunes may be just cool enough to condense both H$_2$O and NH$_3$. However, in this work we neglect the condensation of NH$_3$ at 2 AU, for better comparison with the 2 AU Jupiter models that only condense H$_2$O. The 5 AU and 10 AU Neptune models have both H$_2$O and NH$_3$ clouds. 

 Our $1\times$ Jupiter models appear to be a bit cooler than those shown in \citet{sud05}. This is probably due to our computing a global mean temperature profile which assumes that the absorbed incident flux is redistributed by atmospheric dynamics evenly around the planet.  \citet{sud05} instead compute a mean dayside temperature profile without energy redistribution.  While the efficiency of energy redistribution is an interesting topic for the very hot Jupiters, at the lower effective temperatures relevant to this study the radiative time constants are long and the incident flux is likely to be evenly distributed, as it is on the solar system jovian planets, which do not exhibit any significant diurnal thermal differences.

\subsection{Albedo Spectra Model} \label{subs3_2} 

Given the atmosphere models generated by the radiative-convective model we next apply the albedo spectra model. This albedo spectra model has previously been used for a variety of planetary objects, and is described in detail in \citet{too77, too89, mck89} and \citet{mar99a, mar99b}. In this section we describe the configuration of the albedo spectra model used here. The albedo model takes as input the exoplanet model's gravity and atmospheric temperature, pressure, composition, and cloud properties that are generated within the radiative-convective model discussed in \S \ref{subs3_1}. 

The albedo spectra model uses the same chemistry and opacity database as were used for obtaining the thermal structure.  In the thermal structure model, opacities were binned into intervals using k-coefficients.  For the albedo calculation we use monochromatic opacities over the $\sim790$ points from 0.35 $\mu$m to 1 $\mu$m.  The opacity of course varies with altitude through the model and we account for the changing gaseous composition as well as the influence of temperature and pressure on the opacities themselves.  In this work, we do not model photochemistry or attempt to account for photochemical hazes or chromophores (see \citet{wes04} for a review), which are responsible for the low albedos for the real Jupiter in the blue and UV ($< 0.4$ $\mu$m) and also for its ruddy complexion; this is discussed in more detail in \citet{mar99a}.  While we expect that photochemistry will play an important role in exoplanet spectra, we reserve such considerations for future work.

\subsubsection{Radiative transfer description} \label{subs3_2b}

\begin{figure}
\epsscale{0.5}
\plotone{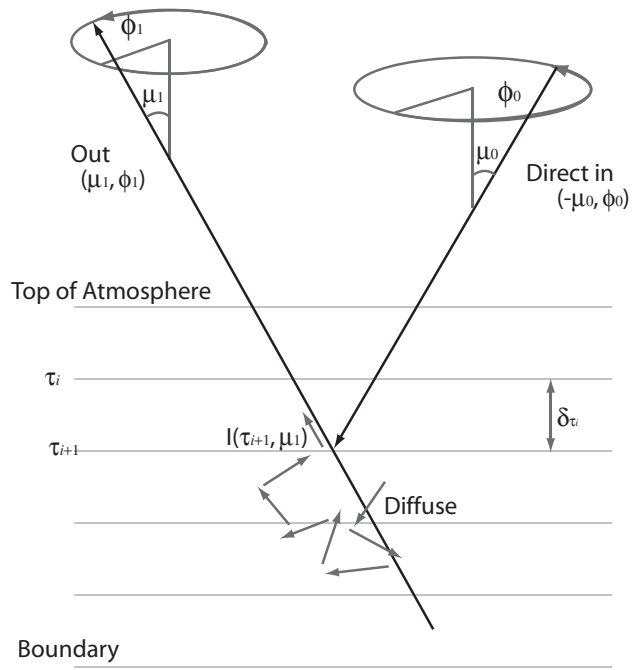}
\caption{Diagram of the plane-parallel geometry for the albedo spectra model, where $\mu_0$ represents the cosine of the angle of incidence, $\beta$, and $\mu_1$ represents the cosine of the angle of reflection, $\vartheta$. \label{planegeom}}
\end{figure}

Figure \ref{planegeom} shows a schematic of the plane-parallel model geometry with incident angle $\mu_0 = \cos\beta = \sin\eta\cos\left(\zeta-\alpha\right)$ and observed angle $\mu_1 = \cos\vartheta = \sin\eta\cos\zeta$, where $\zeta$ and $\eta$ are planetary coordinates of longitude and co-latitude, respectively. 

We use a finite plane-parallel atmosphere divided into $N$ levels and $N$-1 layers. The plane-parallel framework is shown in Figure \ref{planegeom}. Emergent intensity is $I\left(\tau, \mu_1\right)$ with optical depth $\tau_i$ from level $i$ to the top of the atmosphere and incremental change in optical depth of $\delta_{\tau_i} = \tau_{i+1}-\tau_i$ between levels. Both incident and emergent angles are positive from the surface normal, with incident light from direction $\mu_0= \cos\beta$ and the observer direction $\mu_1 = \cos\vartheta$. A negative sign is used to denote the direction of incident flux, e.g., $-\mu_0$. With a source at $\mu_0$, the exact expression for the emergent intensities through the $N$ levels is:

\begin{equation}
I\left(\tau_i, \mu_1, \phi_1 \right) = I\left(\tau_{i+1}, \mu_1, \phi_1\right)e^{-\delta_{\tau_i}/\mu_1} + \int_{0}^{\delta_{\tau_i}} S\left(\tau', \mu_1, \phi_1\right)e^{-\tau'/\mu_1}\frac{d\tau'}{\mu_1} 
\label{eq_a00}
\end{equation}

\noindent where $S\left(\tau', \mu_1, \phi_1\right)$ is the source function and contains contributions from both direct and diffuse scattering. The incident collimated flux crosses the unit area perpendicular to direction defined by $-\mu_0$ and $\tilde{\omega}$ is the `single-scattering' albedo, a parameter used to represent the amount of scattering taking place versus absorption. $\tilde{\omega}$ is calculated for each layer and at each wavelength by taking the ratio of the scattering coefficients to the total scattering and absorption coefficients in a layer. The asymmetry parameter, $g$, is similarly calculated for each layer and wavelength using asymmetry contributions and opacities of the layer constituents. The scattering and asymmetry contributions include an approximation to Raman scattering, Rayleigh scattering and clouds in this model.  

While we begin with an exact solution to the radiative transfer in Eq. (8), we introduce an approximation by substituting a two-stream solution for the scattered radiation field (intensity) in the source function.  In the case of negligible diffuse radiation (when single scattering dominates, as it tends to for clear atmospheres) this approximation has no effect.  As the importance of multiple scattering increases (e.g., when there are clouds and hazes), the influence of this approximation grows.  We present our two stream solution with some detail, but stress that the albedo calculation as used here is {\em not} simply a `two stream solution' \citep{too89}.


The radiative transfer equation is often solved using approaches that do not preserve the angular dependence of the scattering phase function, such as the two-stream methods summarized in \citet{mea80} or the Feautrier method in \citet{mih78}. Here we follow the `source function' approach in \citet{too89} where we first use two-stream quadrature to solve for the diffuse scattered radiation field.  We use the resulting two-stream intensity $I_{t}$ as an approximation to the source function ($S_v$ versus $S_{vt}$ in Eq. 1 and Eq. 52 of \citet{too89}). We are then able to approximate the angular dependence $\left(\mu,\phi\right)$ and $\left(\mu',\phi'\right)$ in the scattering phase function. We use the two-stream `source function' approach here largely because of its compatibility with the existing model. In a future paper we will compare the approach in this work with other radiative transfer methods that preserve the angular dependence of the scattering phase function, such as the $\delta-$M stream method \citep{wis77}. Comparisons with models that include the vector nature of polarized scattering (e.g. \citet{sta06, bue09}) instead of the scalar approach taken for speed and simplicity in this work, would also be interesting.


Following the two-stream source function method \citep{too89}, we use two-stream quadrature (sometimes called discrete ordinates) to solve the radiative transfer equation for the upward and downward diffuse fluxes. The two-stream quadrature approach is only one of many different two-stream methods to approximate the diffuse fluxes. Other approaches, including the hemispheric constant, Eddington, modified Eddington, modified quadrature, and delta-function methods, are treated in detail in sources such as \citet{mea80}, \citet{lio80}, and \citet{too89}. In the form of the `unified description' of two-stream methods in \citet{mea80}, the Gaussian quadrature parameters $\gamma_i$ used in this work are:

\begin{equation}
\gamma_1 =  \frac{\sqrt{3}}{2}\left[2-\tilde{\omega}\left(1+g\right)\right]
\label{eq_a20}
\end{equation}

\begin{equation}
\gamma_2 = \frac{\sqrt{3\tilde{\omega}}}{2}\left(1-g\right)
\label{eq_a21}
\end{equation}

\begin{equation}
\gamma_3  =  \frac{1}{2}\left(1-\sqrt{3}g\mu_0\right)
\label{eq_a22}
\end{equation}

\begin{equation}
\gamma_4 = 1-\gamma_3 = \frac{1}{2}\left(1+\sqrt{3}g\mu_0\right)
\label{eq_a23}
\end{equation}

The forward-scattering nature of many atmospheric particulates, especially clouds, is not well captured by the two-stream approach as described above. The delta-function technique takes advantage of the similarity principle to split out a fractional forward scattering contribution $f$ from original scattering phase function into a delta function, and adjust the asymmetry factor $g$, single scattering albedo $\tilde{\omega}$, and optical depth $\tau$ accordingly. This concept is treated in detail in \citet{jos76} who use it along with an Eddington two-stream approach. We use a delta-function approach as described in \citet{lio80} along with the two-stream source function method in this work as well.  


\subsubsection{Boundary conditions} \label{subs3_2d}

We use the recursive Eq.~\ref{eq_a00} to calculate the level intensities from the `ground', $N$ to the top of the atmosphere. Note that in discussion of boundary conditions it is common to refer to the `ground' or `surface'. For gas giant exoplanets, these terms are not accurate since there is no `ground' in the sense that there is one for a terrestrial planet. Instead, we treat the `surface' as a pressure level deep enough that no photons are reflected back out. We set the following boundary condition for the ground level:

\begin{equation}
I\left(\tau_N,\mu_1\right) = R\left(\mu_1\right)\frac{F^{\downarrow total}\left(\tau_N\right)}{\pi}
\label{eq_a01}
\end{equation}

\noindent where $F^{\downarrow total}\left(\tau_N\right) = F^{\downarrow}\left(\tau_N\right) + F^{\downarrow direct}\left(\tau_N\right)$ is the total downward flux onto the surface, both diffuse and direct. We treat the ground layer as a Lambert surface with reflectance $R\left(\mu_1\right) = R$. In this work, we use a `black' surface, $R = 0$. However, in the general case where $R \neq 0$, we would need to solve the diffuse radiative transfer equation to get $F^{\downarrow}\left(\tau_N\right)$. 

We also set boundary conditions at the top of the atmosphere: there is no diffuse flux coming from outside the atmosphere, and any upward flux consists of the reflected direct flux off the Lambert surface plus the downward diffuse flux scattered upward by the Lambert surface:

\begin{equation}
F_{top}^{\downarrow}  = 0 \mbox{  and  } F_{Surface}^{\uparrow} = R\left(F_{Surface}^{\downarrow direct} + F_{Surface}^{\downarrow}\right)
\label{eq_a02}
\end{equation}

These are the boundary conditions used with the two-stream source function method described above.  

\subsubsection{Scattering} \label{subs3_2c}

The albedo spectra model uses a wavelength-dependent approximation of anisotropic scattering. The cloud model, which uses a full Mie scattering model \citep{ack01}, currently returns the single scattering albedo $\tilde{\omega}$ and asymmetry parameter $g$ as a function of wavelength for each condensate species and at each temperature and pressure level. In future work, we will update the cloud model to return the angular distribution of the scattering in addition to the asymmetry parameter. With this type of approach we would then use an M-term fit to the full angular distribution of the scattering function returned by the cloud model in conjunction with a $\delta-$M-stream radiative transfer approach. For the moment, we start with the available $\tilde{\omega}$ and $g$ (which are functions of wavelength, temperature, and pressure) as input parameters to a two-term Henyey Greenstein function as an approximation to the anisotropic scattering (as discussed below, the two-term HG function approximates the angular dependence of both forward and backward scattering). As with many previous studies, for our current goal of exploring a wide range of exoplanet atmosphere models that may be observed by generally photon-starved direct imaging observations, our treatment of atmospheric scattering is of a scalar, not vector, nature, meaning that the effects of polarization are not considered. We refer readers interested in the effects of polarization to simulations of Jupiter-like and Rayleigh scattering atmospheres in \citet{sta04, sta05, sta06, bue09}.  

We also approximate Raman and Rayleigh scattering. The approximate treatment of Raman scattering is the same as that as described in \citet{pol86}, incorporating Raman scattering into the single scattering albedo as a function of the stellar spectrum (see Eq. 2 in \citet{pol86}). We approximate Rayleigh scattering by including an additional term in the scattering phase function as briefly described below. 

The scattering phase function $p\left(\cos\Theta\right)$ is a representative function that takes contributions from a number of different scatterers in the atmosphere into account. The scattering phase functions used in this model attempt to represent the cumulative scattering effect of all of the constituents in each layer of the model. As noted earlier, the single scattering albedo $\tilde{\omega}$ and asymmetry factor $g$ used in the scattering phase function are updated by the cloud model and radiative-convective model at each layer and for each wavelength with weighted contributions from the atmospheric constituents. 

The scattering angle $\Theta$ can also be written in terms radiation incoming from $\left(\mu', \phi'\right)$ that is scattered into $\left(\mu,\phi\right)$:

\begin{equation}
\cos\Theta = \mu\mu' + \left(1-\mu^2\right)^{1/2}\left(1-\mu'^2\right)^{1/2}\cos(\phi'-\phi)
\label{eq_a7}
\end{equation}

The second term on the right hand side is sometimes neglected in the case of azimuthal independence or averaging. The scattering phase function may be written in terms of $\cos\Theta$ or $\mu',\phi'$ and $\mu,\phi$. The integral of the scattering phase function is typically normalized either to unity or to the single scattering albedo, $\tilde{\omega}$.  Note that the scattering phase function $p$ should not be confused with planet phase function $\Phi\left(\lambda, \alpha\right)$. Also note the $\pi$ difference in directionality between the scattering angle and the planet's phase angle, $\alpha = \pi-\Theta$; backscattering would be more strongly seen by an observer at full phase, $\alpha = 0^{\circ}$, whereas backscattered $\Theta = 180^{\circ}$.

As shown in Fig.~\ref{jup3xtau}, for these models, the contribution of scattering to the albedo spectra not only depends on the mixture of gases in a given layer, but also the presence, composition, location, and depth of clouds in the atmosphere. There are many different analytic scattering phase functions that attempt to realistically capture the scattering behavior of atmospheric constituents, such as the Henyey-Greenstein function, as well as approaches to capturing strong forward scattering using delta functions.  While there are a variety of scattering phase functions that are used to represent  different types of atmospheric constituents, it is common to expand the phase function in terms of Legendre polynomials in $\cos\Theta$. 

\begin{equation}
p\left(\cos\Theta\right) = \sum_{\ell = 0}^{N}\tilde{\omega_{\ell}}P_{\ell}\left(\cos\Theta\right)
\label{eq_a5}
\end{equation}

\begin{equation}
\tilde{\omega_{\ell}} = \frac{2\ell+1}{2} \int_{-1}^{1} p\left(\cos\Theta\right)P_{\ell}\left(\cos\Theta\right)
\label{eq_a6}
\end{equation}

\noindent Note that despite the similar notation, the single scattering albedo, $\tilde{\omega}$, is different from the Legendre constants $\tilde{\omega}_{\ell}$. In most two-stream models, the Legendre polynomial expansion of the scattering phase function is not used beyond the first order, $P_1$. In that case $P_0\left(\cos\Theta\right) = 1$ and $P_1\left(\cos\Theta\right) = \cos\Theta$. For a second order expansion, $P_2\left(\cos\Theta\right) = \frac{1}{2}\left(3\cos^2\Theta-1\right)$. 

For a first order expansion, azimuthal independence is often assumed, leading to:

\begin{equation}
p\left(\mu,\mu'\right) = 1 + 3g\mu\mu'
\label{eq_a9}
\end{equation}

\noindent where $g$ is the asymmetry factor, defined as the first moment of the phase function, and also equal to $\tilde{\omega}_{1}/3$. The asymmetry factor $g$ represents the relative strength of forward scattering. 

The Rayleigh scattering phase function is: 

\begin{equation}
p\left(\cos\Theta\right) = \frac{3}{4}\left(1+\cos^2\Theta\right)
\label{eq_a37}
\end{equation}

\noindent The $\cos^2\Theta$ dependence is not included in the first-order expansion in Eq.~\ref{eq_a9}. However, continuing the Legendre expansion to second order allows representation of Rayleigh scattering with the $\cos^2\Theta$ term:

\begin{equation}
p\left(\cos\Theta\right) = \tilde{\omega}_0 + \tilde{\omega}_1\cos\Theta + \frac{\tilde{\omega}_2}{2}\left(3\cos^2\Theta-1\right)
\label{eq_a38}
\end{equation}

For both isotropic scattering and Rayleigh scattering, $g = 0$.  We use the original first order expansion (Eq.~\ref{eq_a9}) for $p$ in the initial radiative transfer calculation of the two-stream fluxes that are to be used in the source function method; the truncation treats the Rayleigh scattering as isotropic (since for Rayleigh, $g = 0$). However, once we have used the source-function method to plug the fluxes back in, we use a second-order phase function expansion for $p$ and assume azimuthal symmetry such that:

\begin{equation}
p\left(\mu\mu'\right) = 1 + 3g\mu\mu' + \frac{g_2}{2}\left[3\mu^2\left(\mu'\right)^2-1\right]
\label{eq_a39}
\end{equation}

\noindent With this expansion, Rayleigh scattering can be modeled when $g = 0$ and $g_2 = \frac{1}{2}$. In the model, $g$ goes to zero when the contribution from $\tau_{Rayleigh}$ in the denominator is large, and $g_2$ goes to $1/2$. For the purpose of evaluating the second-order expansion with the two-stream source function, we also define a `mean square angle' $\bar{\mu}_2$:

\begin{equation}
\bar{\mu}_2^2 = \frac{\int \mu^2 I d\mu}{\int I d\mu}
\label{eq_a41}
\end{equation}

\noindent where $\bar{\mu}_2$ is treated as a free parameter ranging from 0 to 1 that depends on the two-stream angle approximation used, and that, in combination with the higher order term $g_2$, is intended to provide an accurate result for the pure Rayleigh case. In the model used, $\bar{\mu}_2 \approx \left(1/\sqrt{3}\right)^{1/2}$. The weighted asymmetry parameters ($g, g_2$) in the model balance the relative contribution of Rayleigh scatterers such that the diffuse scattering phase function approaches a Rayleigh scattering phase function when Rayleigh scattering dominates. This capability becomes more important at short wavelengths and in clear (cloud and haze-free) atmospheres, and is particularly relevant for the direct imaging of exoplanets in the optical. 

\subsubsection{Two term Henyey Greenstein} \label{subs3_2e}

\begin{figure}
\epsscale{1}
\plotone{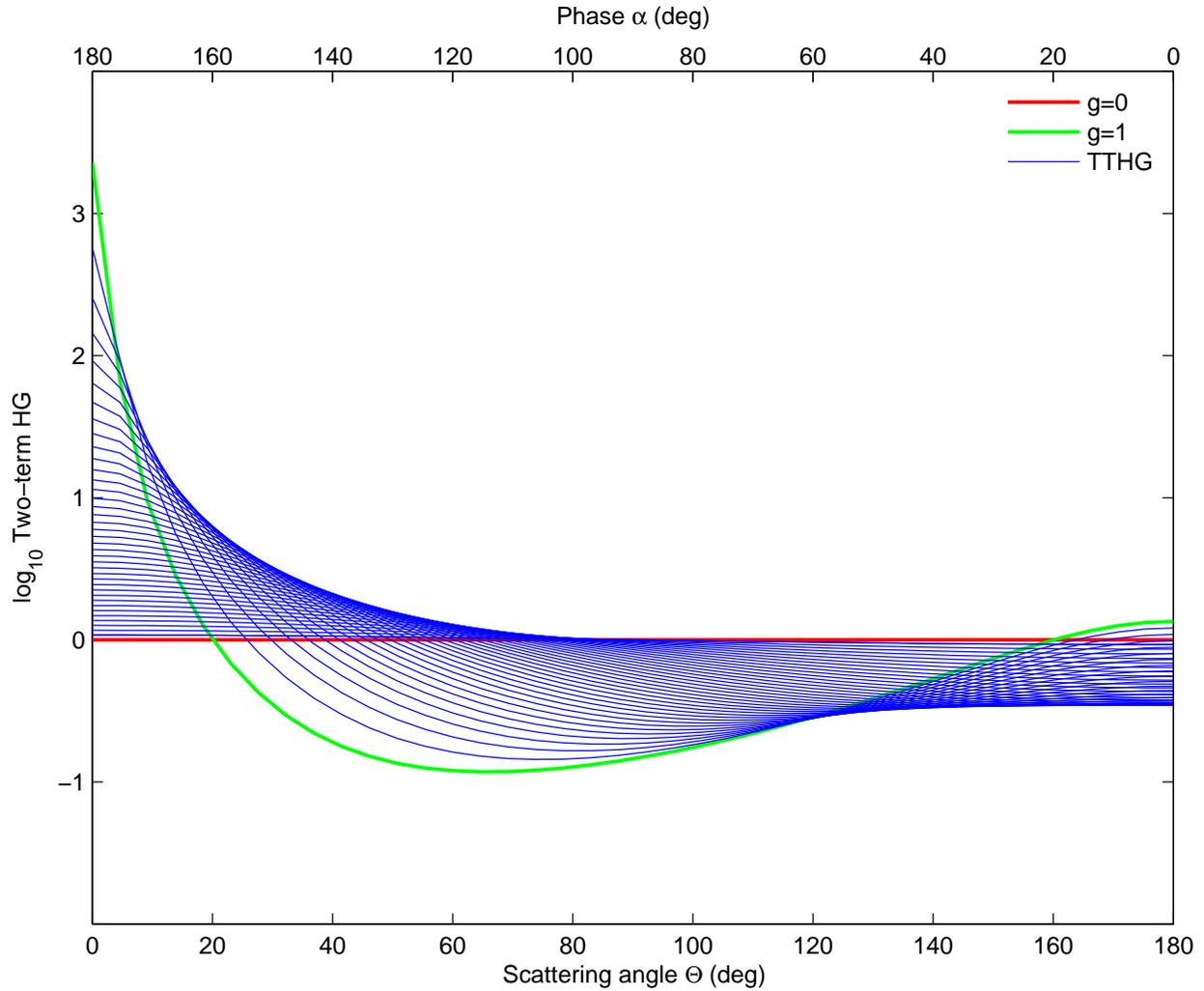}
\caption{The two-term Henyey-Greenstein phase function used in this work, shown with respect to both the scattering angle as well as phase angle $\alpha$ for reference. Backscattering is at $\Theta = 180^{\circ}$ ($\alpha = 0^{\circ}$) and forward scattering at $\Theta = 0^{\circ}$ ($\alpha = 180^{\circ}$). \label{tthg}}
\end{figure}

We use a two term Henyey-Greenstein function (TTHG) as the direct scattering phase function to capture the contributions of very strong forward scattering and moderate back scattering. The TTHG makes use of the asymmetry parameter $g$ that is calculated in the radiative-convective model for the clouds that form within each layer following \citet{ack01}. As noted above, we plan to update this approach to retain the angular properties from each layer and use an M-term Legendre polynomial to best fit the angular distribution of the scattering phase function for each layer. For the moment, we use a TTHG that is a function of $g$ for each layer and that has a form generally consistent with high forward scattering and moderate back scattering by clouds and aerosols, as shown in Fig.~\ref{tthg} and informed by studies of the scattering function in solar system giant planets in \citet{pol86, dyu05} and discussion of the scattering properties of clouds and aerosols in \citet{lio80}. The values of $g$ from the cloud model used in this work, at pressure levels where clouds have formed, typically range between 0.5 at short wavelengths to 0.99 at long wavelengths.

\noindent The general form of a single Henyey-Greenstein function is: 

\begin{equation}
p_{\rm HG} = \frac{1-g^2}{\left(1 + g^2 -2g\cos\Theta\right)^{3/2}}
\label{eq_s71}
\end{equation}

\noindent The HG phase function is useful because it both approximates strongly peaked forward scattering and has a simple expansion in Legendre polynomials:

\begin{equation}
p_{\rm HG} = \sum_{\ell = 0}^{N}\left(2\ell + 1\right)g^{\ell}P_{\ell}\left(\cos\Theta\right)
\label{eq_s72}
\end{equation}

\noindent From its expansion, $p_{\rm HG}$ has a first moment of $g$, and has a second moment $f = g^2$. The ability to expand the HG function in terms of $N$ Legendre polynomials as well as the simple form of its second moment also makes it useful for adjusting two-stream solutions to better include the strong forward scattering nature of cloud particles and aerosols \citep{jos76}. However, a single term HG scattering phase function does not capture backscattering. To capture backscattering, two term HG scattering phase functions can be used:

\begin{equation}
p_{\rm TTHG} = b p_{\rm HG}\left(g_a,\Theta \right) + \left(1-b\right)p_{\rm HG}\left(g_b,\Theta\right)  
\label{eq_s73}
\end{equation}

\noindent where $g_a > 0$ causes the forward peaks and $g_b < 0$ the backward peaks. The form used here has $g_a = g$, $g_b = -g/2$, and $b = 1-g_b^2$.  


\subsubsection{Planet Phase Geometry} \label{subs3_3}
 
To compute the albedo spectra, we integrate over emergent intensities resulting from the incident radiation. Previously \citep{mck89, mar99a}, the emergent intensity calculations assumed an $\alpha = 0^{\circ}$ system geometry where the source and observer were collinear, such that $\mu_0 = \mu_1$. With that restriction, to consider variation with phase, the geometric albedo spectra could then be multiplied by some planet phase function $\Phi\left(\lambda,\alpha\right)$ to emulate the change in emergent flux with phase. Such an approach does not capture the changes in the albedo spectra with wavelength that occur for different phases as the paths through the atmosphere and contributions from anisotropic scatterers change. We thus have updated the albedo spectra model to use different $\mu_0$ and $\mu_1$.

\begin{figure}
\epsscale{0.5}
\plotone{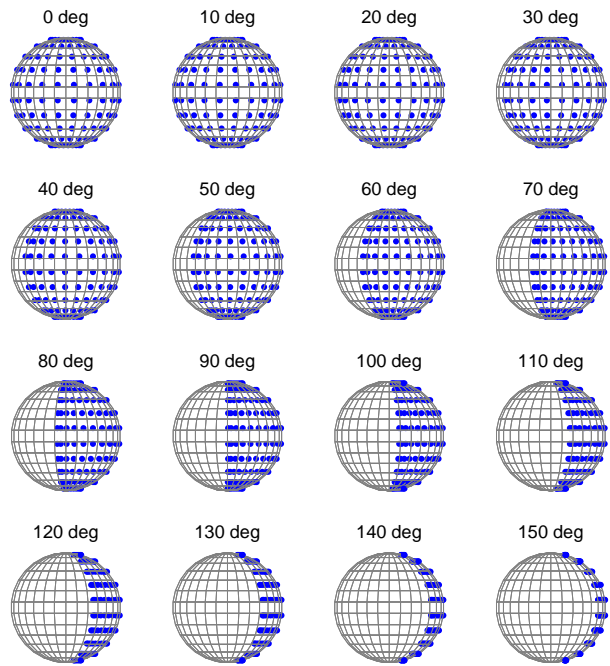}
\caption{Each dot represents one plane-parallel albedo spectra model run with a different pair of incident $\mu_{0}$ and emergent $\mu_{1}$ angles. A two-dimensional Chebyshev-Gauss integration over all dots is performed to calculate the albedo spectra as a function of planet phase. There are $n = 100$ dots shown per planet phase. \label{diskdots}}
\end{figure}

To simulate a spherical planet, we cover the illuminated surface of a sphere with many plane-parallel facets, where each facet has different incident and observed angles, as shown in Figure \ref{diskdots}. Following the approach of \citet{hor50,hor65}, we use two-dimensional planetary coordinates and Chebyshev-Gauss integration to integrate over the emergent intensities and calculate the albedo spectra. The Chebyshev-Gauss angles are the input angles to the intensity calculations. The plane-parallel approach currently does not accurately capture curved geometry (limb effects) at high phase angles,  e.g. \citet{cha31a, cha31b}; we will include this refinement in a future update.





\section{Results} \label{sec4}

In this section, we present the  albedo spectra model and place our results in context with previous work. We first examine the cloud structures as a function of planet-star separation and consider their effect on the optical depth $\tau$. Next, we present the albedo spectra for all of the exoplanet models at full phase ($\alpha = 0^{\circ}$). Then we examine how the albedo spectra change as $\alpha$ varies from $0^{\circ}$ to $180^{\circ}$ for representative cases.
\subsection{Clouds and Planet-Star Separation} \label{subs4_0}

\begin{figure}
\epsscale{1}
\plotone{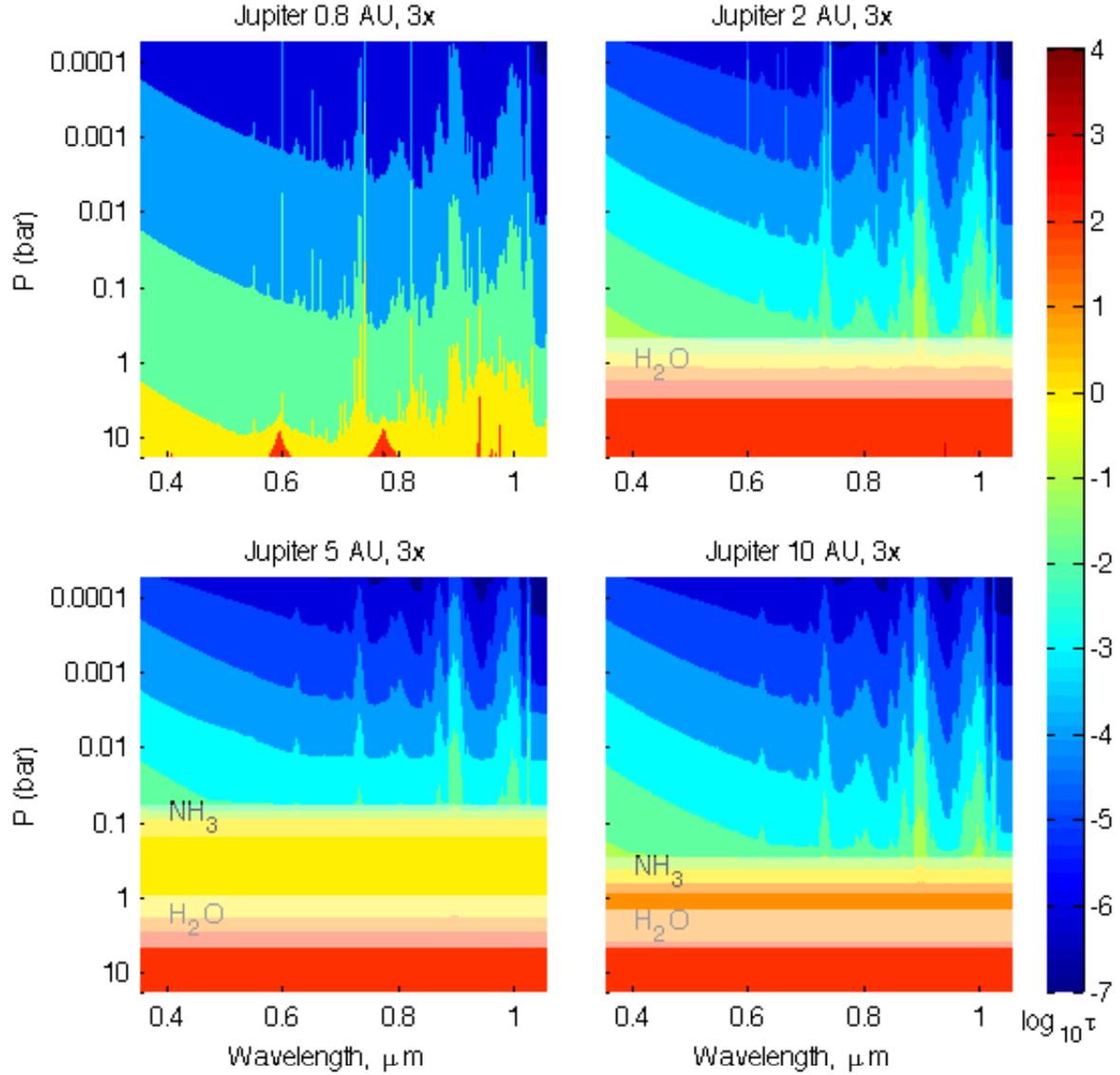}
\caption{These plots show how the optical depth, $\tau$, varies as a function of pressure and wavelength for the Jupiter 3$\times$ models. The background shading represents $\log\tau$, with values corresponding to the color scale on the right. Optical depth of 1 is at $\log\tau = 0$. The H$_2$O and NH$_3$ cloud structures represent the location of the clouds in the models. See \S \ref{subs4_0}. \label{jup3xtau}}
\end{figure}

\begin{figure}
\epsscale{1}
\plotone{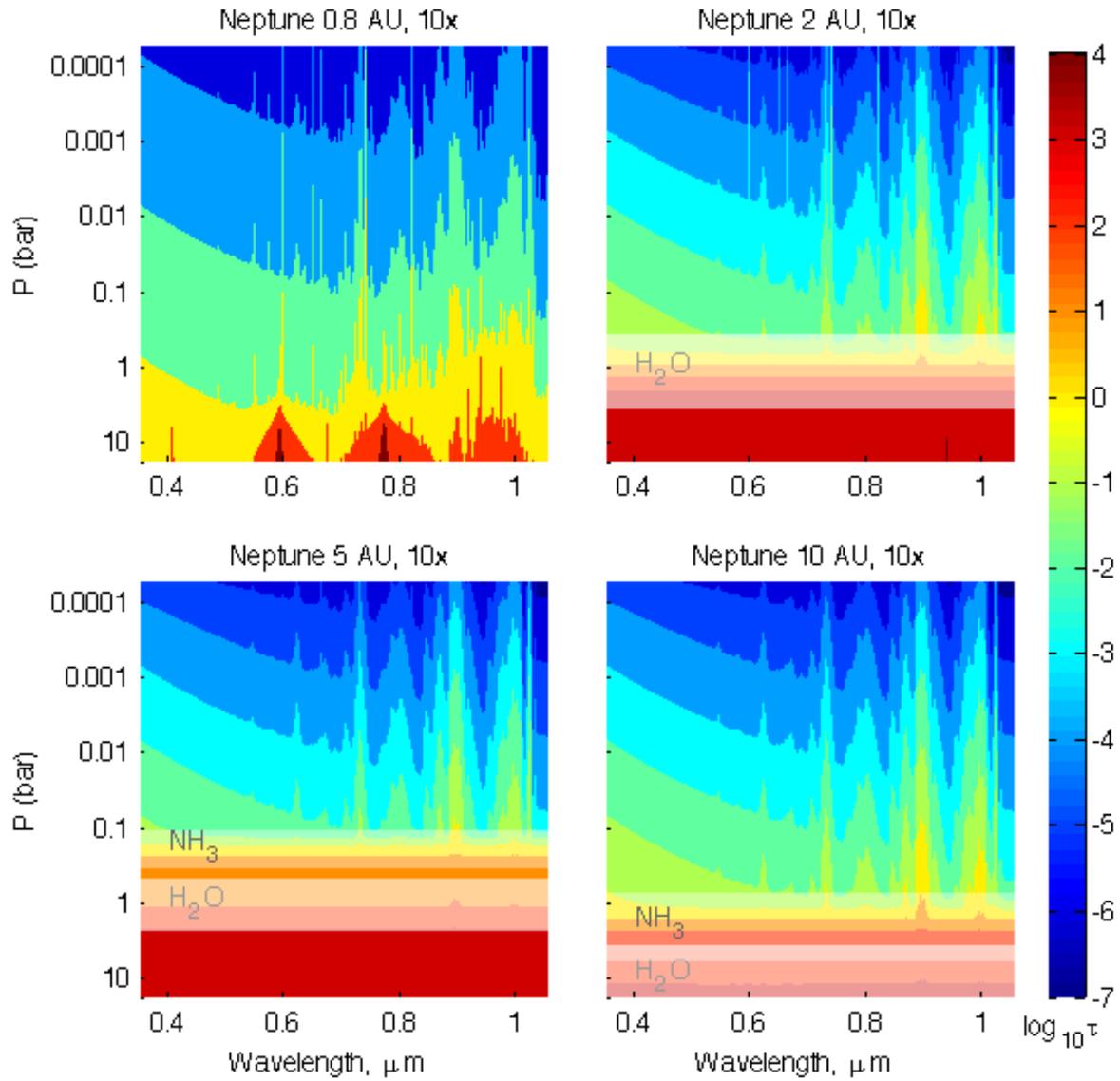}
\caption{As in Figure \ref{jup3xtau} but for the Neptune 10$\times$ models. \label{nep10xtau}}
\end{figure}

\begin{figure}
\epsscale{1}
 \plottwo{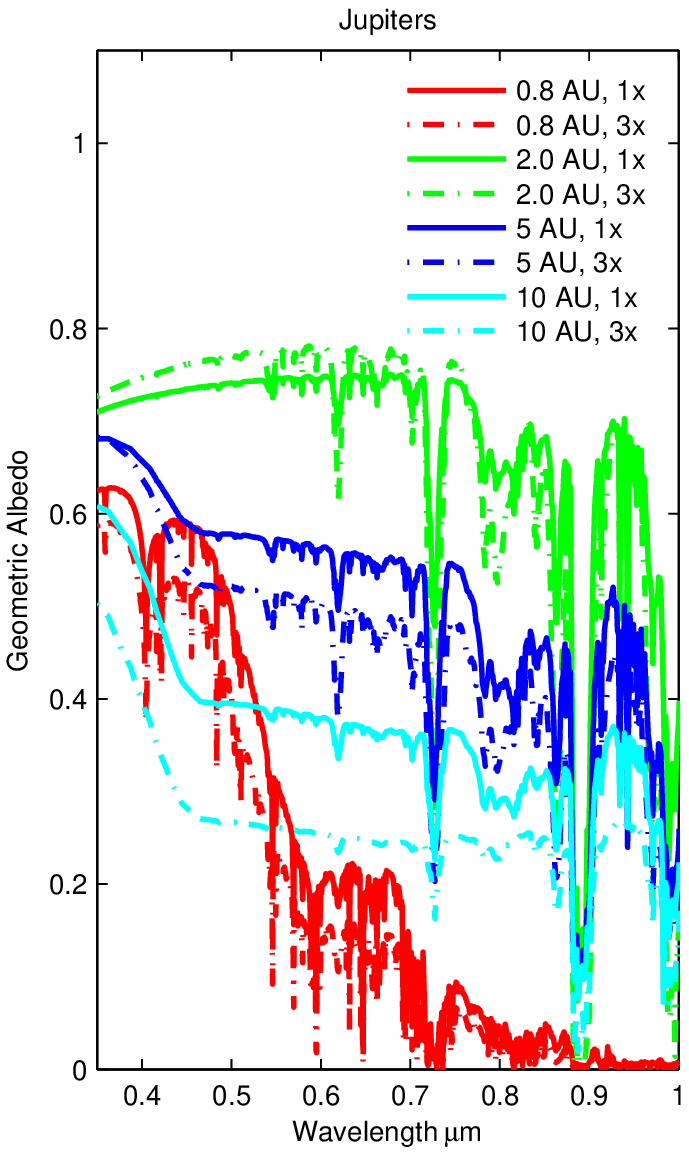}{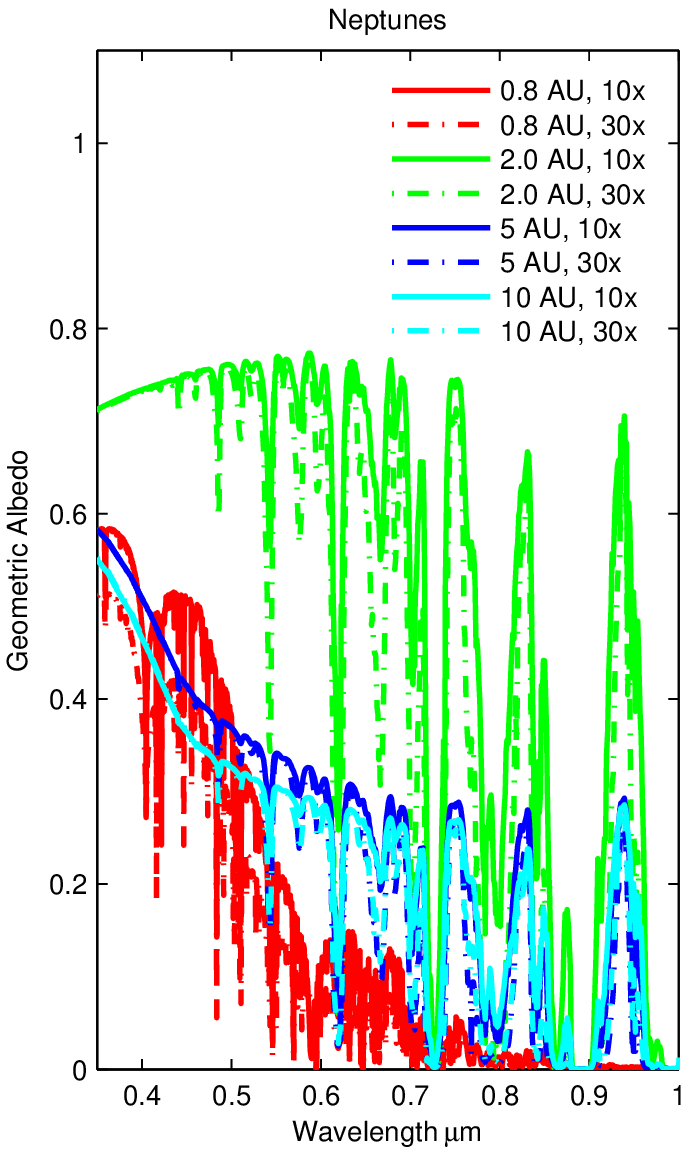}
\caption{Geometric albedo spectra ($\alpha = 0^{\circ}$) for each of the exoplanet model atmospheres used in this work. The models cover a range of planet-star separations from 0.8 AU to 10 AU, and a range of heavy-element abundances (metallicities) with respect to solar ($1\times$). The Jupiter models have $1\times$ (solid) and $3\times$ (dashed) solar abundance, and the Neptune models have $10\times$ (solid) and $30\times$ (dashed) solar abundance. See \S \ref{subs4_0} and \ref{subs4_1}.
\label{metspectra}}
\end{figure}

In Figures \ref{jup3xtau} and \ref{nep10xtau}, we show how the optical depth $\tau$ varies throughout the model atmospheres with increasing planet-star separation for our $3\times$ Jupiter and $10\times$ Neptune models.  The corresponding albedo spectra are shown in Fig.~\ref{metspectra}.  Figures \ref{jup3xtau} and \ref{nept10xtau} shows pressure $P$ vs. wavelength with shaded contours of $\log\tau$ that correspond to the color scale on the right. Optical depth of $\tau = 1$ is at $\log\tau = 0$. 

For the cloud-free $3\times$ Jupiter case at 0.8 AU, Fig.~\ref{jup3xtau} shows that Rayleigh scattering dominates the short wavelengths deep into the atmosphere. Contributions from the alkalis, Na ($\sim0.59$ $\mu$m) and K ($\sim0.78$ $\mu$m, see Table \ref{tab2}) shape the spectrum and are evident deeper in the atmosphere where they are favored by chemical equilibrium.  Thus even though the effective temperature of this model (Table \ref{tab1}) is $290\,\rm K$, well below the temperature at which we would expect the atmosphere to be devoid of alkali elements, alkalis remain an important opacity source.  The reason for this is that the clear atmosphere allows photons to penetrate as deep as 10 bars before they are scattered for the 0.8 AU $3\times$ case (Figure \ref{jup3xtau}, top left panel).  At ten bars the temperature exceeds 800 K and the alkalis are present in the gas and thus are detectable.

At 2 AU, bright and optically thick water clouds form below $\sim500$ mbar; their presence dominates opacity across the optical. Mie scattering predicts that the extinction efficiency at optical wavelengths will be essentially constant for all particles with radii greater than $1\,\rm \mu m$ and thus the spectral contribution of the clouds appears flat in the figures. It is interesting to note that optical thicknesses can reach $\sim10,000$ for H$_2$O clouds. At 5 AU, ammonia clouds form below $\sim50$ mbar and water clouds form below $\sim1$ bar, at a deeper level than the for the warmer 2 AU case. The ammonia clouds also contribute significantly to the optical depth. For the still colder 10 AU Jupiter $3\times$ model, both the NH$_3$ and H$_2$O clouds form deeper into the atmosphere, allowing Rayleigh scattering to again dominate the opacities at short wavelengths. Even at long wavelengths, unit optical depth is likely to be reached before encountering the clouds. 

In addition to the alkalis (clearly seen at 0.8 AU) and CH$_4$, opacity due to gaseous H$_2$O plays a role between $\sim0.92$ and 0.95 $\mu$m in our albedo spectra for all of the cases closer in than 5 AU.  Detection of water absorption features in the optical have not been confirmed in Jupiter's reflection spectrum, although absorption features near $\sim0.94$ $\mu$m were noted by \citet{kar94} as being present. At the time, H$_2$O was suggested potentially being the cause, however, later observations suggested that NH$_3$ could be responsible \citep{kar98}. We address the presence of H$_2$O again later in discussion of Figure \ref{karcomp}.

Figure ~\ref{nep10xtau} shows the relationship between $\tau$, $P$, $\lambda$ and planet-star separation for the $10\times$ Neptune models. Similar to the 0.8 AU $3\times$ Jupiter case, the 0.8 AU $10\times$ Neptune case is dominated by Rayleigh scattering at short wavelengths, only with more pronounced absorption features into the red. The 2 AU case is also dominated by the presence of H$_2$O clouds that form just below $\sim380$ mbar, a bit higher than the Jupiter case, implying that the albedo spectra should be a bit brighter as well. At 5 AU, an ammonia cloud forms below $\sim100$ mbar, and extends down nearly to the H$_2$O cloud, which forms below $\sim470$ mbar. Also similar to the 10 AU $3\times$ Jupiter case, the 10 AU $10\times$ Neptune shows both cloud decks forming at substantially higher pressures, below $\sim700$ mbar for NH$_3$ and below $\sim3.6$ bar for H$_2$O.

\subsection{Geometric Albedo Spectra ($\alpha=0^{\circ}$)} \label{subs4_1}

\begin{figure}
\epsscale{1}
 \plotone{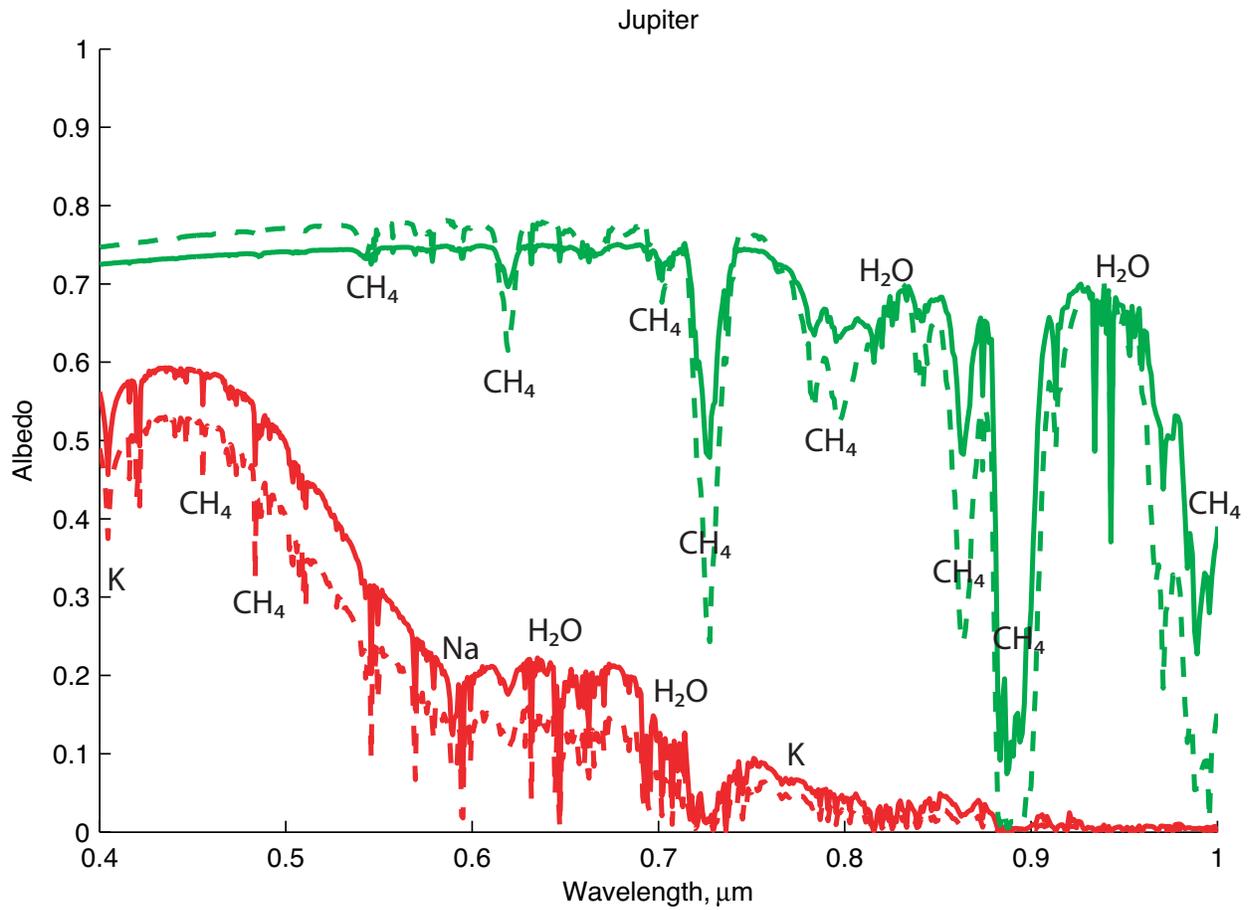}
\caption{Geometric albedo spectra of Jupiter-analogs at 0.8 AU (red) and 2 AU (green) and $1\times$ (solid) and $3\times$ (dashed) solar heavy element abundances; prominent spectral features are noted: CH$_4$, K, Na, and H$_2$O.  See Table \ref{tab2}. \label{jupfeatures}}
\end{figure}

In Figure \ref{metspectra} we show albedo spectra for all of the model cases summarized in Table \ref{tab1}. The spectra for our Jupiter-like exoplanet models can be compared with those presented in \citet{mar99a,sud00} and \citet{sud05}. For the purpose of comparison, the model Jupiters used here map to the Class III (clear), Class II (water cloud) and Class I (ammonia cloud) nomenclature used in \citet{sud00}. Our clear and ammonia cloud models are similar to those in \citet{sud00}, and our water cloud models at 2 AU are a bit brighter. As shown in Fig.~\ref{jupfeatures}, distinct Na, K, CH$_4$ and H$_2$O features are apparent, particularly CH$_4$ near 0.62, 0.74, and 0.89 $\mu$m (see Table \ref{tab2}). As discussed in \S \ref{subs4_0}, the cloud-free 0.8 AU spectra are dominated by Rayleigh scattering at short wavelengths for both $1\times$ and $3\times$ Jupiters and $10\times$ and $30\times$ Neptunes. 

At any given planet-star separation, the higher metallicity compositions generally have smaller albedos due to the increased opacity of their atmospheres. The presence of relatively high and thick H$_2$O clouds in the atmosphere of the 2 AU Jupiters result in a higher albedo across the visible. The bright effect of relatively high H$_2$O clouds is also apparent through mid-band in the 2 AU Neptunes, however, the absorption features are considerably more pronounced at longer wavelengths for the Neptunes than for the Jupiters. At full phase, backscattering from clouds also plays a role. At separations of 5 AU and 10 AU, the albedos show progressively larger contributions by Rayleigh scattering at short wavelengths, consistent with the lower clouds in Figs.~\ref{jup3xtau} and \ref{nep10xtau}. At 5 AU and 10 AU, the difference between albedos for $1\times$ and $3\times$ solar abundances of heavy elements for Jupiters is  larger at all wavelengths than the difference between albedos for $10\times$ and $30\times$ solar for Neptunes. In the cooler atmospheres at larger planet-star separations the albedo spectra seem to change little with increasing heavy element abundances above about a ten-fold enhancement.

\begin{figure}
\epsscale{1}
 \plottwo{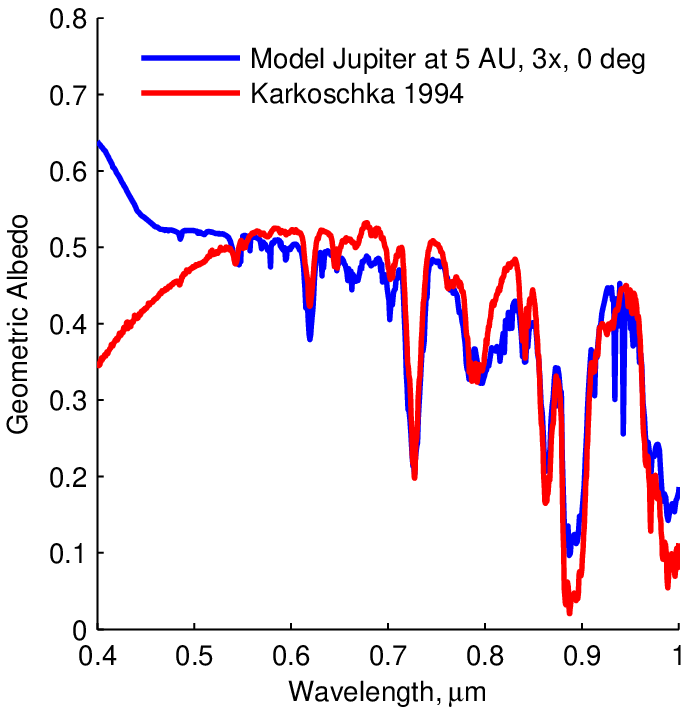}{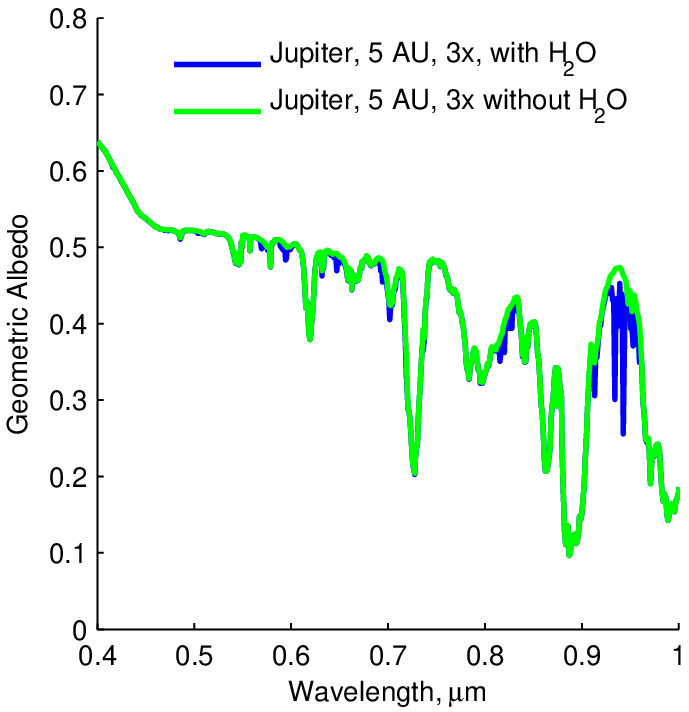}
\caption{Left, comparison of albedo spectra from our `untuned' 5 AU Jupiter-mass exoplanet model ($3\times$ solar abundance of heavy elements) with data of Jupiter obtained by \citet{kar94}. See \S \ref{subs4_1}. Right, identification of the features near 0.94 $\mu$m in our model: they are water features. \label{karcomp}}
\end{figure}

We compare the albedo spectra for our standard 5 AU $3\times$ enhancement Jupiter model at $\alpha = 0^{\circ}$ with observed data from the real Jupiter in our solar system at near full-phase from \citet{kar94} in Fig.~\ref{karcomp}.  This figure illustrates how our interpretation of Jupiter might proceed if we were to detect it as an exoplanet.   The general agreement in morphology of the spectral features is simply a consequence of the spectrum being primarily (but not exclusively) shaped by methane.   As noted in \S \ref{subs3_2}, we do not include the effect of photochemical products such as hazes that would explain the difference between our model and the observed data at short wavelengths \citep{mar99a, sud00}. Although our 5 AU $3\times$ Jupiter model was not adjusted to fit the data, varying the cloud thickness or $f_{\rm sed}$ parameter would brighten or darken the spectra. 

The difference between our 5 AU $3\times$ Jupiter model and the data from \citet{kar94} around 0.94 $\mu$m are of interest.  As noted above \citet{kar94} and \citet{kar98} noticed features near this wavelength region in Jupiter's albedo spectrum. \citet{kar94} mentioned that they could potentially be water features, but in the later paper concluded they were more likely to be ammonia. While the features we see in this region of our models are likely not the same as those observed by \citet{kar94} and \citet{kar98}, we were able to confirm, by running models both with and without H$_2$O as shown in Fig.\ref{karcomp}, that the model features around 0.94 $\mu$m are indeed attributable to H$_2$O. These features would be interesting targets for direct imaging observations of exoplanets since the appearance of water is a sensitive probe of temperature and composition. 


\subsection{Albedo Spectra vs. Phase} \label{subs4_2}

\begin{figure}
\epsscale{1}
 \plottwo{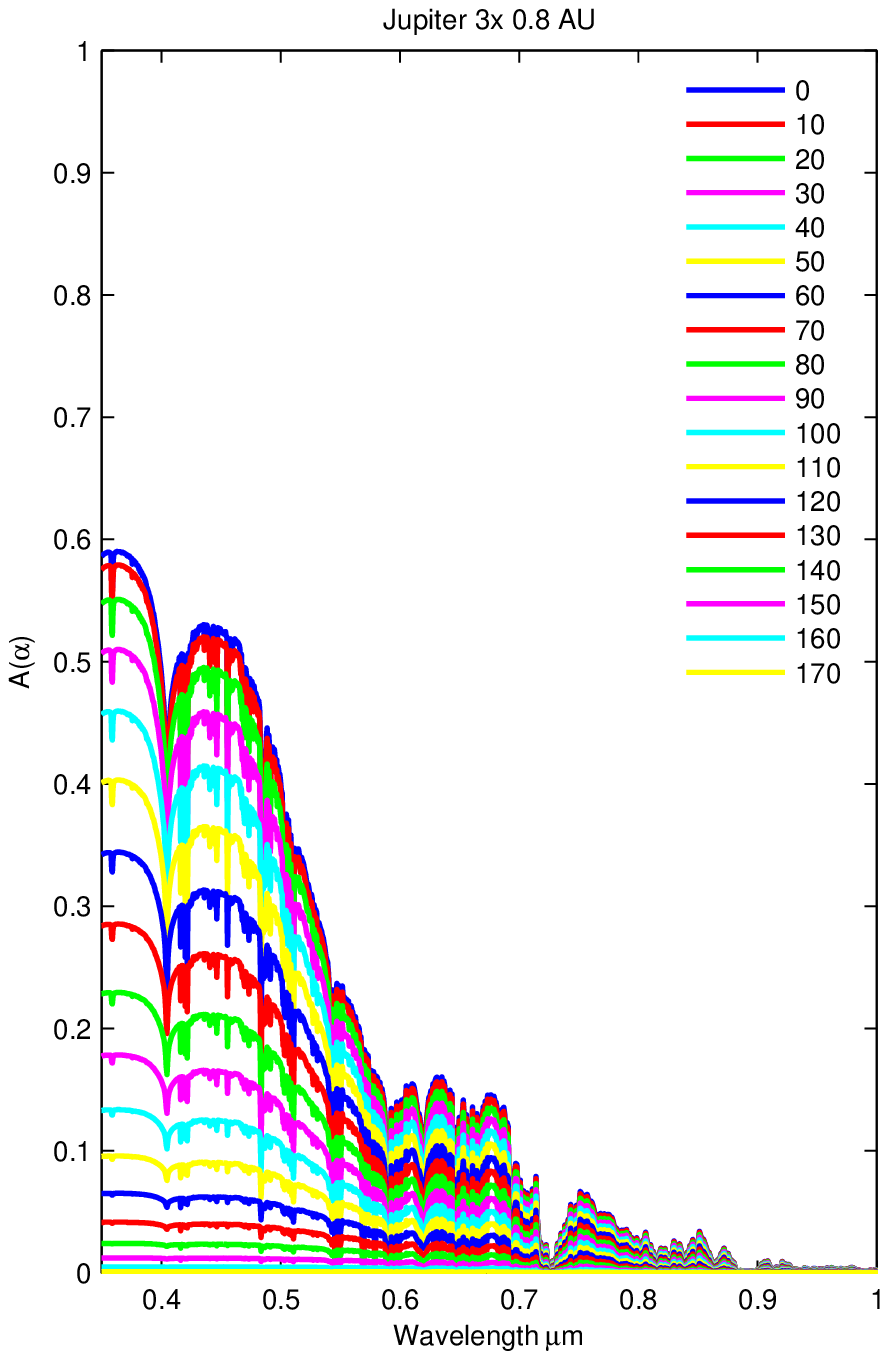}{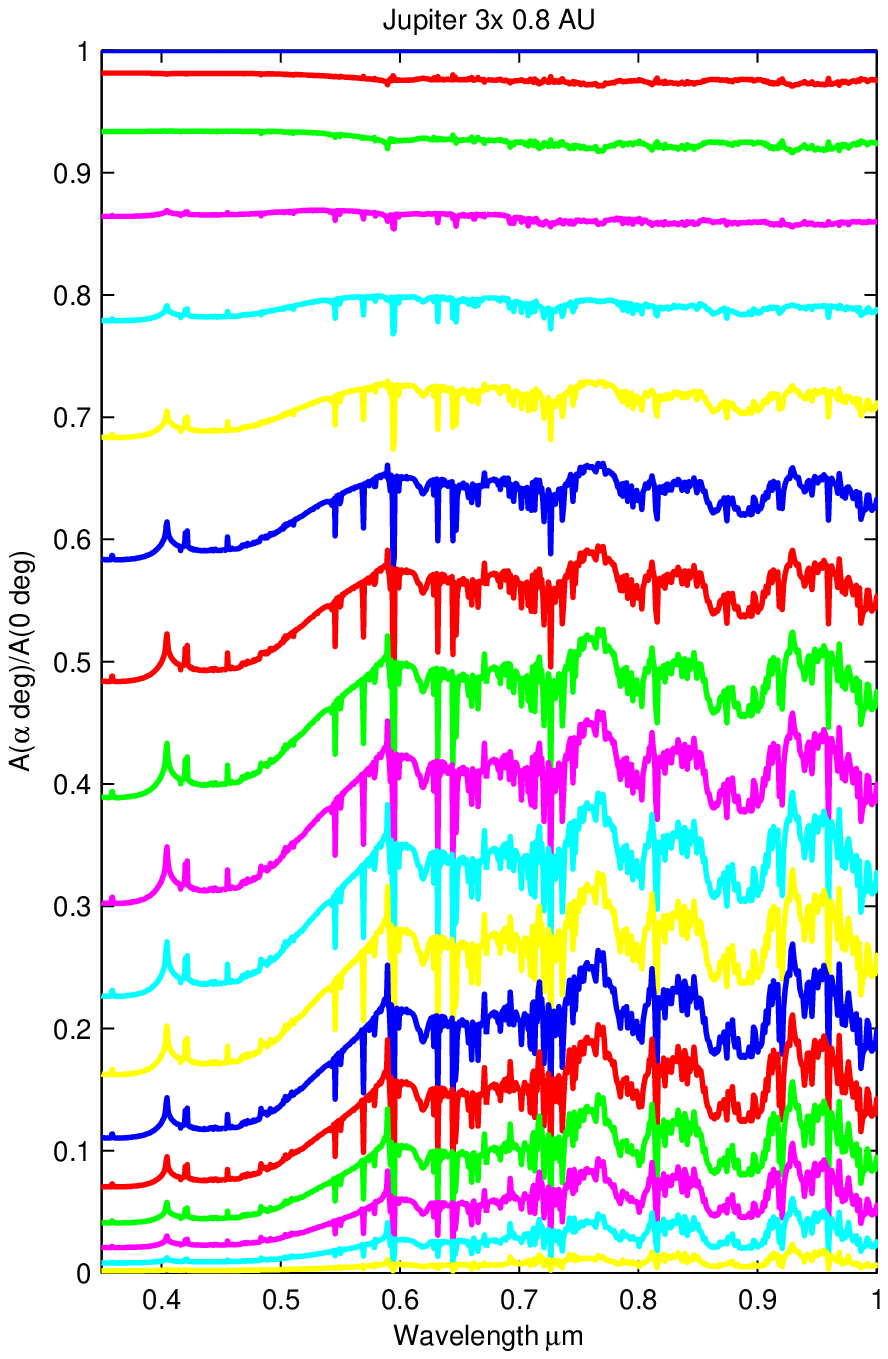}
\caption{Albedo spectra of the 0.8 AU $3\times$ Jupiter as a function of planet phase $\alpha$. Left, albedo spectra vs. $\alpha$. Right, ratio of albedo spectra in increments of $\alpha = 10^{\circ}$ to that of the albedo spectrum at $\alpha=0^{\circ}$. Note the variation in the ratio as a function of wavelength; see \S \ref{subs4_2}. \label{jupphase}}
\end{figure}

In Fig.~\ref{jupphase} we show the albedo spectra of the 0.8 AU $3\times$ Jupiter model as $\alpha$ progresses from $0^{\circ}$ (full) to $180^{\circ}$ (new). As expected, the albedo decreases with increasing $\alpha$. More importantly, Fig.~\ref{jupphase} also shows the ratio of the albedo spectra at each $\alpha$ increment to the geometric albedo spectrum at $\alpha = 0^{\circ}$. The wavelength dependent changes are apparent as a function of $\alpha$. There are both broad effects on the spectrum, such as the drop in brightness from shorter to longer wavelengths becoming shallower as the planet phase increases, as well as narrower effects on the spectrum due to the changing line depths as different volumes of atmosphere are probed at different phases. 

\begin{figure}
\epsscale{1}
 \plottwo{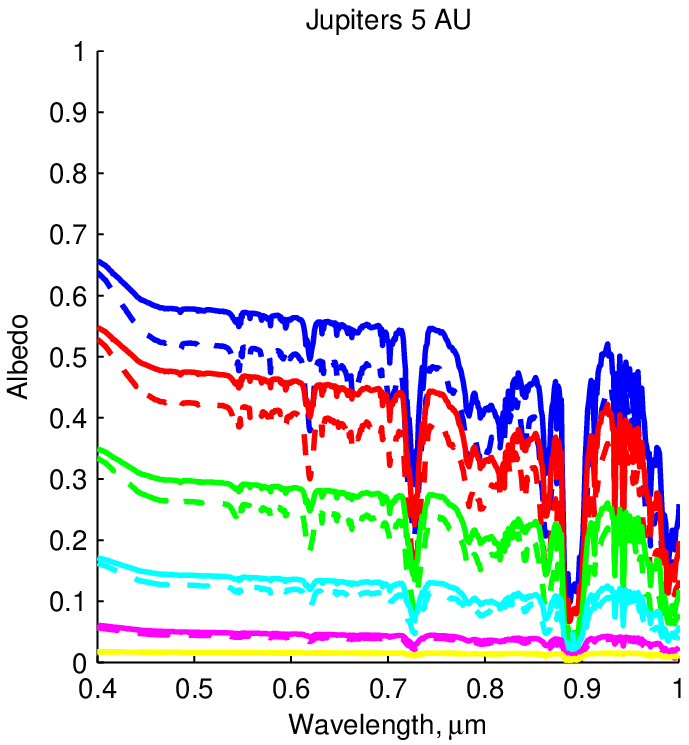}{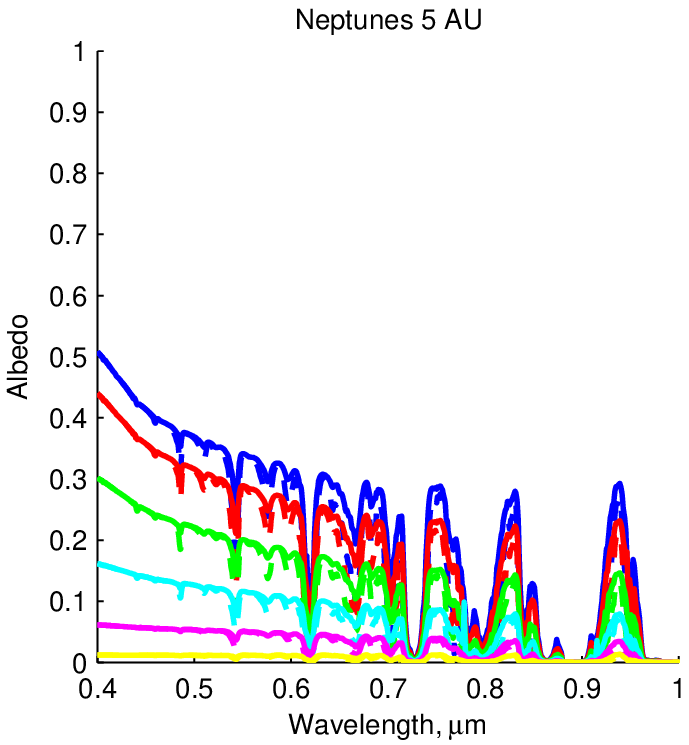}
\caption{The effect of metallicity as a function of $\alpha$ at 5 AU for $\alpha=$ 0$^{\circ}$, 30$^{\circ}$, 60$^{\circ}$, 90$^{\circ}$, 120$^{\circ}$, and $150^{\circ}$ (blue, red, green, cyan, magenta, and yellow, respectively). Left, Jupiters, solid lines are $1\times$ solar, and dashed are $3\times$.  Right, Neptunes, solid lines are $10\times$ solar, and dashed are $30\times$. Note that at 5 AU, the difference in heavy element abundance has a larger effect on the albedo spectra for Jupiters than it does for Neptunes. The $1\times$ Jupiter model is substantially brighter than the $3\times$, likely due to the H$_2$O clouds forming at lower pressure (higher altitude). \label{metphase}}
\end{figure}

Figure \ref{metphase} shows albedo spectra of the Jupiter and Neptune models at 5 AU for a few different phase angles to illustrate the point noted in \S \ref{subs4_1}, that for planet-star separations beyond 2 AU, where it is cooler, there is not as a large of a difference in the albedo spectra for the $10\times$ and $30\times$ Neptune models as there is between the $1\times$ and $3\times$ Jupiter models, even at different phase angles. This implies that it will be more challenging to use albedo spectra to differentiate between Neptune analogs with large abundances of heavy elements. In future work, we plan to further increase the metallicity of Jupiter analogs to determine to what extent this occurs for the Jupiters as well as Neptunes. We also plan to investigate decreasing the metallicity of the Neptune analogs to examine the range of metallicities that we can differentiate between at cooler, larger planet-star separations. For the Jupiter case, it appears that the difference in metallicity is easier to detect when probing deeper into the atmosphere at smaller phase angles than at larger phase angles.


\subsection{Phase Functions} \label{subs4_3}

\begin{figure}
\epsscale{0.8}
\plotone{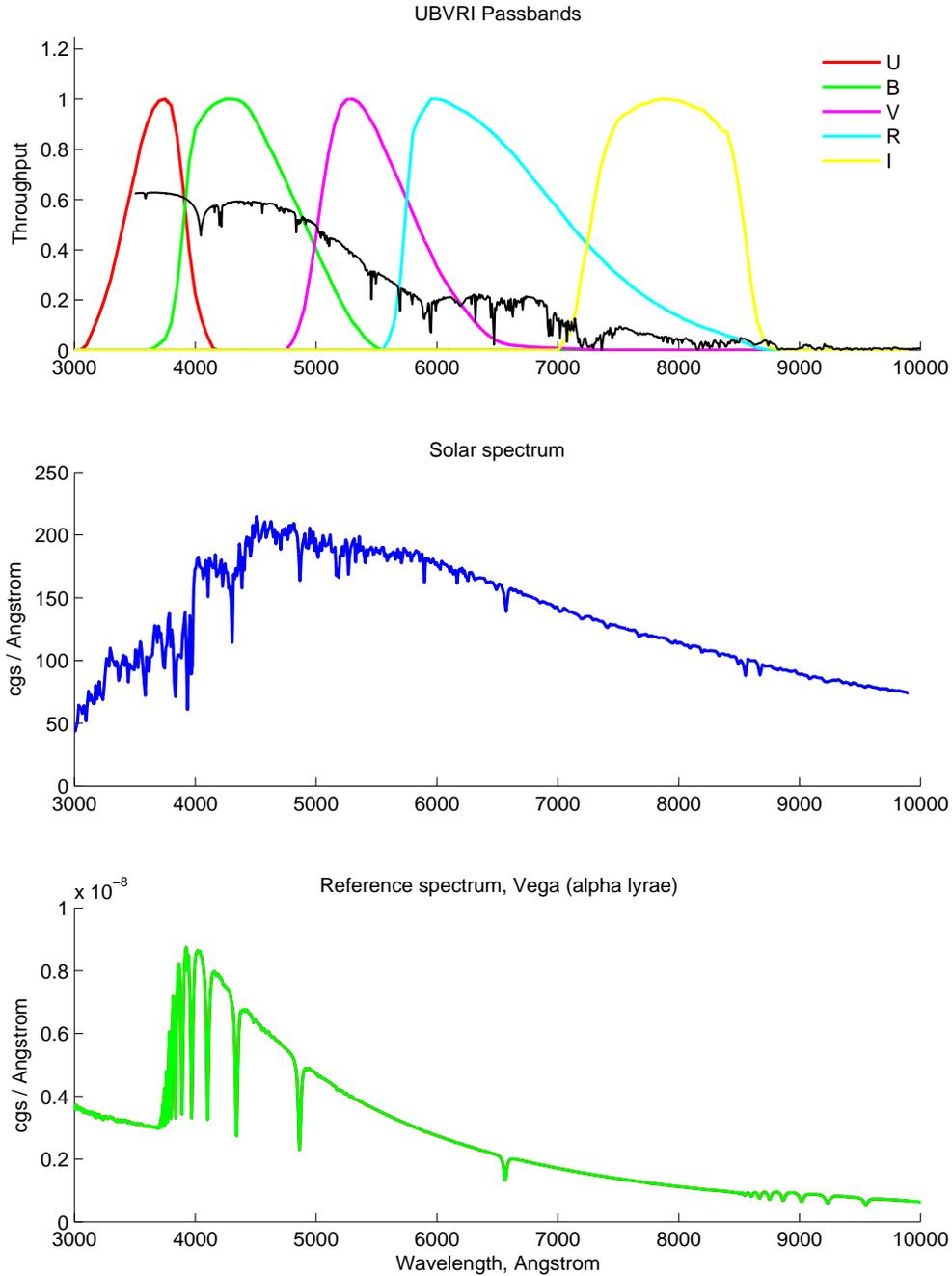}
\caption{Reference spectra and color filters used in the models and for calculating magnitudes and for color-color comparisons. Upper, the Johnson-Morgan/Cousins UBVRI filter passbands courtesy the Virtual Observatory, shown with albedo spectrum of model Jupiter at 0.8 AU at full phase (black line). Middle, the solar spectrum from the Space Telescope Imaging Spectrograph (STIS) courtesy the Space Telescope Science Institute (STScI). Bottom, the Vega reference spectrum (also courtesy STIS/STScI). See \S \ref{subs4_3} and \ref{subs_a9}. \label{ubvrifig}}
\end{figure}

From the albedo spectra as a function of phase, we can also generate phase functions, $\Phi\left(\lambda,\alpha\right)$ as in Eq.~\ref{eq2}. Although we can calculate a phase function for each wavelength individually, we instead present the phase functions for our model exoplanets using the Johnson-Morgan/Cousins UBVRI filter passbands \citep{fug95} as shown at the top of Fig.~\ref{ubvrifig}. Also shown for reference in Fig.~\ref{ubvrifig} with the filter responses\footnote{Filter responses obtained from the Virtual Observatory, \url{http://voservices.net/filter/filterfindadv.aspx}, September 2009} is the albedo spectrum of the 0.8 AU $1\times$ Jupiter model.  Note that our models extend shortward only to 0.35 $\mu$m. Since they do not extend to the shortest wavelengths of the U filter, we do not use U in this work. Fig.~\ref{ubvrifig} also shows the solar spectrum used for the parent star\footnote{Solar spectrum obtained from STScI, \url{ftp://ftp.stsci.edu/cdbs/calspec}, September 2009}, and the reference spectrum\footnote{Reference spectrum obtained from STScI,  \url{ftp://ftp.stsci.edu/cdbs/calspec}, September 2009} used to compute color magnitudes.  Direct imaging observations will obviously be made with different filter responses, possibly tuned specifically to achieve the particular scientific objectives of the observation. However the approach presented here would be applicable to any arbitrary filter set as well (see Appendix \ref{subs_a9}).

In Figure \ref{pollackfig}, we first compare the phase function for our 10 AU $10\times$ Neptune model with data points of both Uranus and Neptune from Voyager 1 presented in \citet{pol86}. The data points and our phase function are shown with a Lambert phase function for reference. While the error bars are relatively large, and there are two Uranus data points that are closer to the Lambert curve than our model, there is general agreement at higher phases. Since these particular Voyager 1 data were taken with a clear filter\footnote{The Voyager vidicon detector was sensitive from 0.28 to $0.64\,\rm \mu m$.}, the model phase function shown is an average of phase functions over all wavelengths. We did not include the phase function for Jupiter used in \citet{dyu05} and \citet{sud05} for comparison since the Pioneer data used in the generation of that curve were not disk-integrated. 

\begin{figure}
\epsscale{0.8}
\plotone{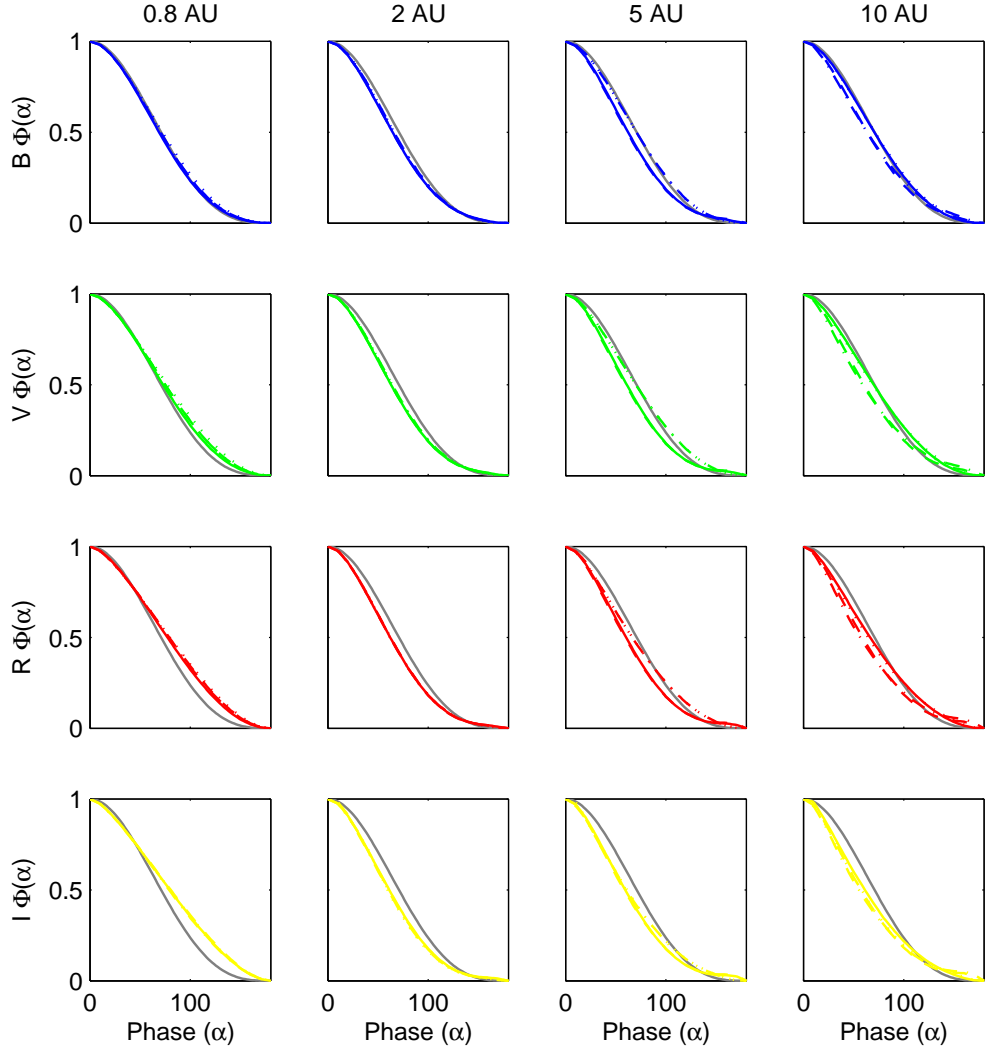}
\caption{Phase functions shown versus planet-star separation and for different color filters (see Fig. \ref{ubvrifig} for filters). Each plot shows the Lambert phase function (gray line) as a reference along with phase functions for the $1\times$ (solid line), $3\times$ (dashed line), $10\times$ (dash-dot), and $30\times$ (dotted) exoplanet models.
\label{phasefunctionfig}}
\end{figure}

Figure \ref{phasefunctionfig} shows phase functions for each model exoplanet and for the B, V, R, and I filters. The phase function is shaped by geometry and the composite of all of the scattering phase functions making contributions to scattering in the exoplanet's atmosphere. For reference, the area beneath a Lambert phase function, or its phase integral $q$, as in Eq.~\ref{eq3}, is $3/2$, and the area beneath a Rayleigh phase function is $4/3$ ($q_{\rm Lambert} > q_{\rm Rayleigh}$). Different line styles represent different metallicities: $1\times$ is solid, $3\times$ is dashed, $10\times$ is dot-dash, and $30\times$ is dotted. A Lambert phase function is shown in each sub-plot for reference. For the 0.8 AU cloud-free case, it is difficult to distinguish between the metallicities, although the higher metallicity phase functions are generally larger than the lower metallicities. The phase functions start off quite similar to the Lambert phase function at short wavelengths, and become shallower and rise above than the Lambert phase function as the colors move redward. 

For the 2 AU water cloud case, the phase functions are all just below the Lambert phase function; the difference becomes more pronounced as colors move redward; for the R and I filters, the beginnings of a forward-scattering `toe' are seen at large phase angles. Effects of backscattering would appear at small phase angles as well. For the 5 AU ammonia and water cloud case, the ammonia clouds are higher in the atmosphere than the water clouds, and there is a noticeable difference between the Jupiters and Neptunes, likely because the ammonia clouds are a bit higher for the Jupiters than they are for the Neptunes (also true at 10 AU, see Figs.~\ref{jup3xtau} and \ref{nep10xtau}). At shorter wavelengths at 5 AU, the Jupiters fall below the Lambert curve and the Neptunes land closer to it, although the Neptune curves also fall below the Lambert curve as colors get redder. The forward-scattering `toe' becomes more pronounced. At 10 AU, both the ammonia and water clouds condense at much higher pressures (lower altitudes), particularly for the Neptunes. The Neptune phase functions are below both the Jupiter phase functions and the Lambert phase function. We note that our phase functions are not nearly as steep near full phase as are those in \citet{sud05}. We will further examine the dependence of phase function morphology on the assumptions made in the treatment of the direct and diffuse scattering functions in a future paper. We discuss in the following section the use of color-color diagrams to help distinguish between the different types of exoplanet models, as it would be difficult to observe and confirm the small differences between the phase functions shown in \ref{phasefunctionfig}.  

From the phase functions and the geometric albedos, we can use the equations outlined in \S \ref{subs2_2} to calculate the phase integral $q$, spherical and Bond albedos $\rm A_s$  and $\rm A_B$. Our Bond albedo calculation integrates over 0.35 to 2.5$\mu$m. Contributions from longer wavelengths are negligible (less than 5\%) for all cases since both the albedo spectra and incident flux drop off quickly at longer wavelengths. Table \ref{tab3} contains geometric albedos calculated for each B, V, R, I color filter for each of our models. Table \ref{tab4} contains the corresponding phase integrals $q$. The Bond albedos  in Table \ref{tab6} yield the effective temperatures in Table \ref{tab1}. 

\subsubsection{Bond albedos and implications for dynamics} \label{subs4_3a}

The Bond albedos shown in Table \ref{tab6} combined with our assumed internal heat flows result in planetary effective temperatures for our model planets shown in Table \ref{tab1} (using equation \ref{eq6}). \citet{ing78} found that if the ratio $E$ of emitted to total absorbed power is {\em greater} than $4/\pi\sim 1.27$ for a giant planet, then internal convection will redistribute heat such that there is vanishing equator to pole temperature gradient, as there is on Jupiter.  Based on our computed Bond albedos and internal energy fluxes (parameterzed as $T_{\rm int}$ in Table 1) then all of our model cases--except for the Jupiters at 5 and 10 AU--are in fact characterized by $E<1.27$.  Thus we expect that in the general case of mature Jupiters and Neptunes (with modest   $T_{\rm int}$) many planets will fall into a regime in which we expect to find substantial equator to pole temperature gradients.  Such planets may be expected to exhibit atmospheric dynamics and cloud patterns quite unlike those expressed by Jupiter and Saturn.


\section{Application to Observations} \label{sec5}

First generation space-based coronagraphic direct imaging observations of widely-separated (cooler) exoplanets will likely consist of photometric measurements using fairly wide filters, and result in coarse spectra or possibly even only few photometric data points in the visible. Here, we use the high-resolution model albedo spectra to generate coarse ($R = 5$ and 15) spectra and color-color diagrams of our gas and ice giant planet models. We compare our model coarse spectra and exoplanet colors to \citet{kar94}'s observations of the giant planets in our solar system.

%
\subsection{Coarse Spectra} \label{subs5_1}

\begin{figure}
\epsscale{1}
 \plottwo{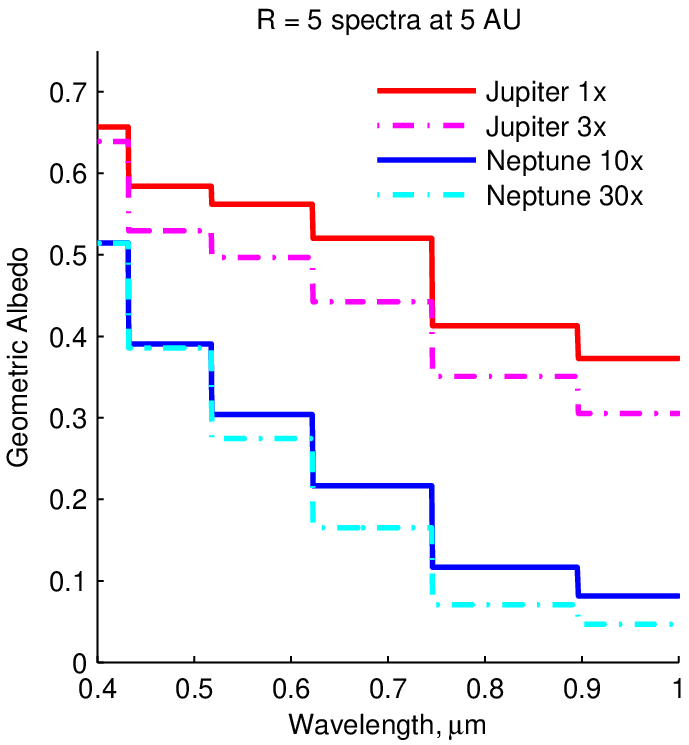}{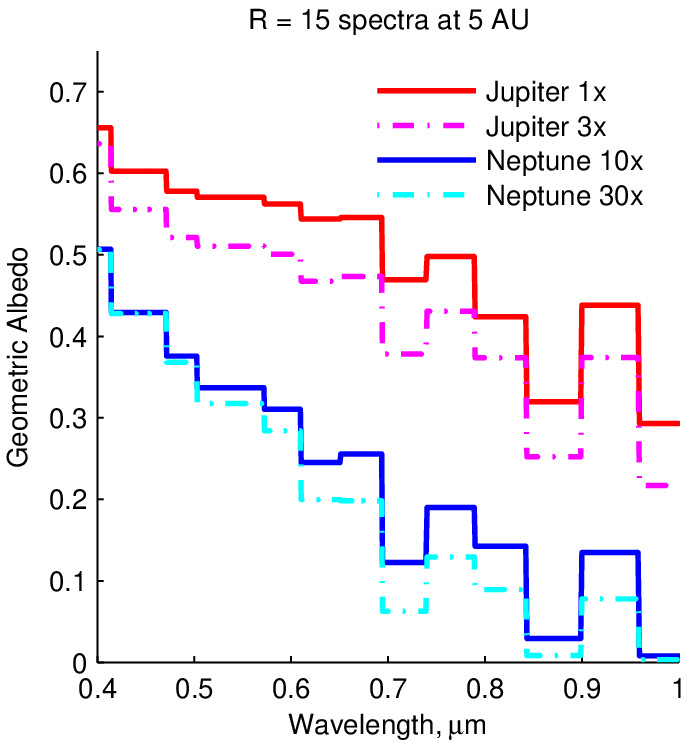}
\caption{Simulated coarse-resolution albedo spectra for Jupiters and Neptunes at 5 AU with different compositions at $\alpha = 0^{\circ}$ for comparison with Figure \ref{lowreskar}. Left, resolution R = 5. Right, resolution R = 15. \label{lowresmodspec}}
\end{figure}

We noted in \S \ref{sec2} that even for the low-resolution spectra of solar system outer planets in Fig.~\ref{lowreskar} generated from \citet{kar94}'s high resolution spectra, it was still possible to discern between gas giant (Jupiter, Saturn) and ice giant (Uranus, Neptune) planets. This is still the case in Fig.~\ref{lowresmodspec}, where the Jupiter models are generally brighter  than the Neptunes. For solar system outer planets,  the real Jupiter and Saturn are darker at short wavelengths due to absorption by photochemical hazes, but are brighter in the middle of the optical. The real Neptune and Uranus are brighter at short wavelengths and darker at longer wavelengths, due to stronger absorption features in the red.  The brightness of a given planet depends sensitively on the height and optical thickness of the uppermost cloud decks, however, so one could imagine planets with somewhat different cloud structures from those presented here.  The actual flux measured by a camera depends on the product of a planet's albedo, cross section, and phase.  Within the limitations of the models (which do not include photochemistry) ice giants generally will be more difficult to detect than gas giants at longer wavelengths both because of their lower albedos and their smaller radii.

Fig.~\ref{lowresmodspec} again illustrates the difference between $1\times$ and $3\times$ solar abundances for the Jupiter models is larger than the difference between $10\times$ and $30\times$ for the Neptune models. In the $R = 15$ spectra in Fig.~\ref{lowresmodspec}, it is also possible to see the effect of the strong methane absorption in all of the spectra between 0.7 and 0.9 $\mu$m. Placement of relatively coarse filters both centered on these features as well as on the continuum around them should make it possible to detect and possibly differentiate between these features in early direct imaging observations.
%
\subsection{Color-Color} \label{subs5_2}

\begin{figure}
\epsscale{0.8}
\plottwo{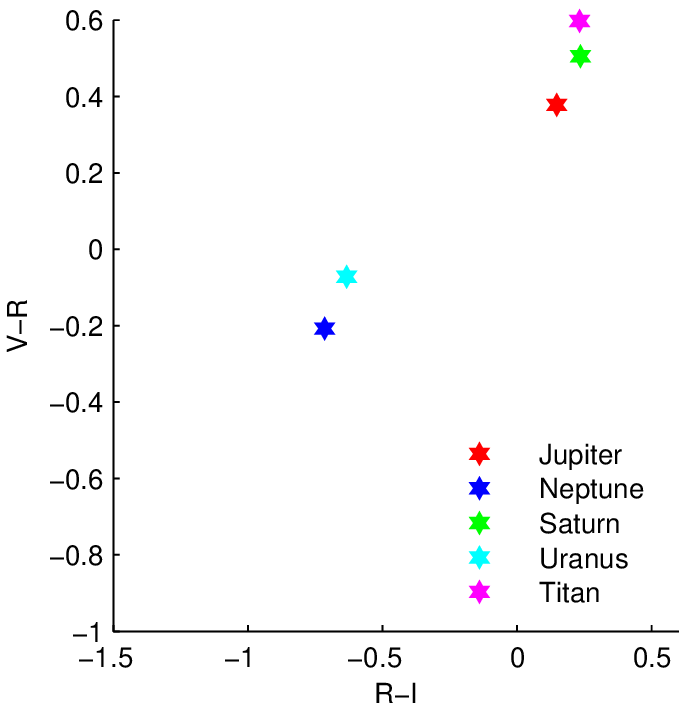}{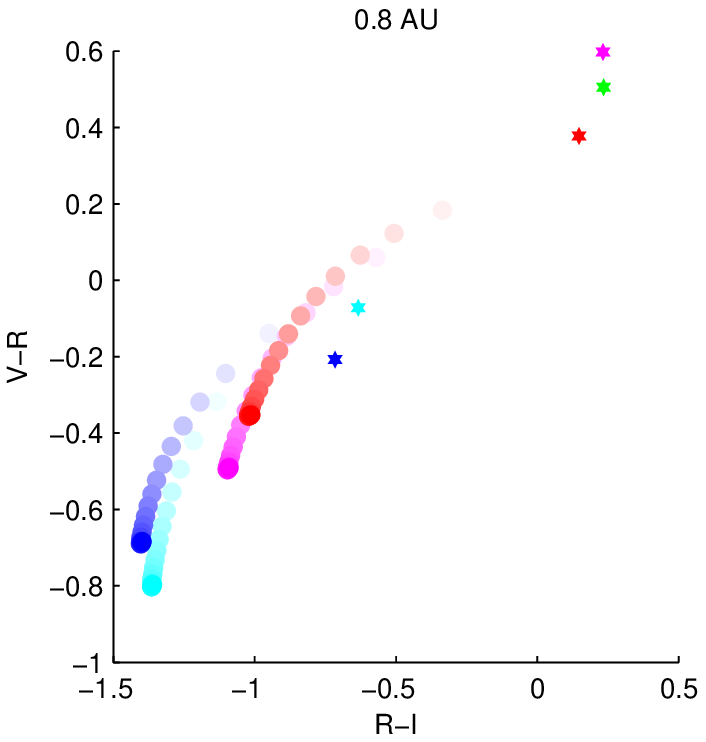}\\
\plottwo{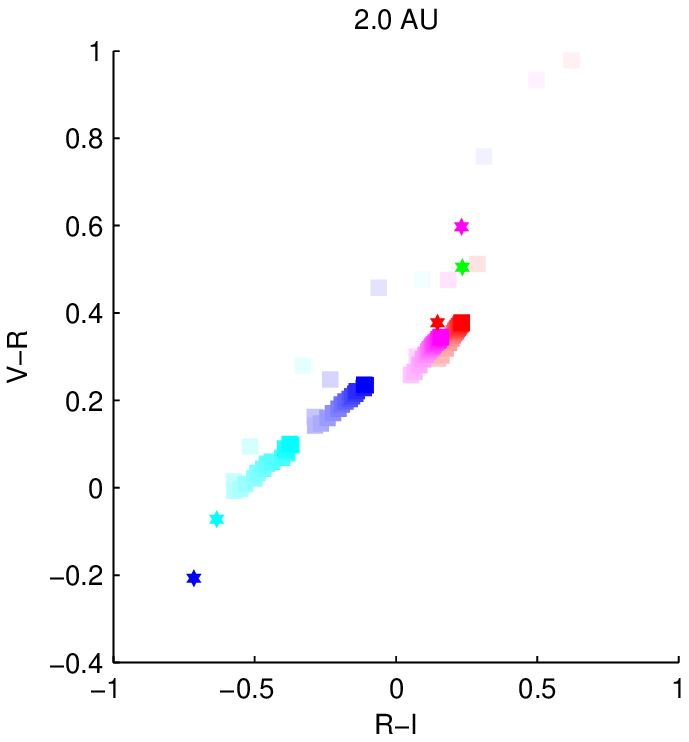}{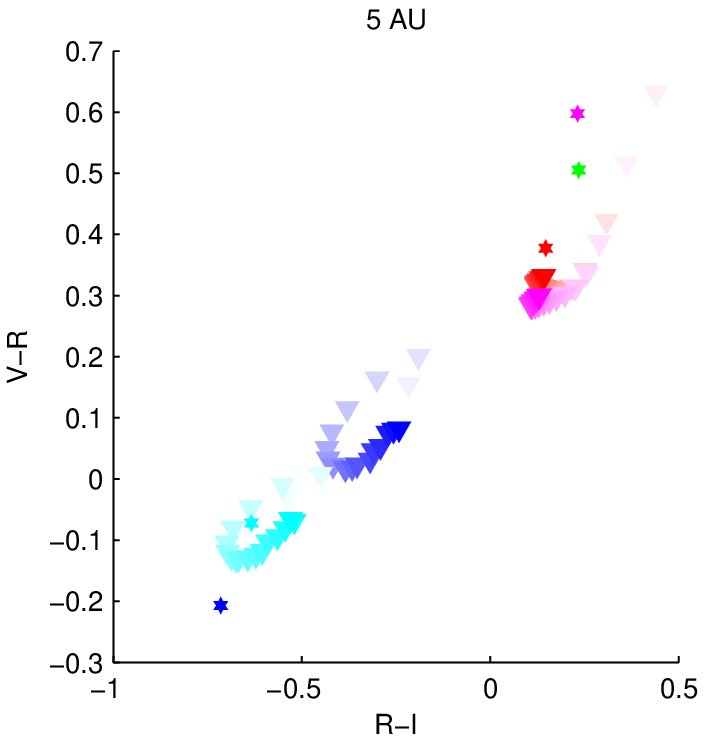}\\
\plottwo{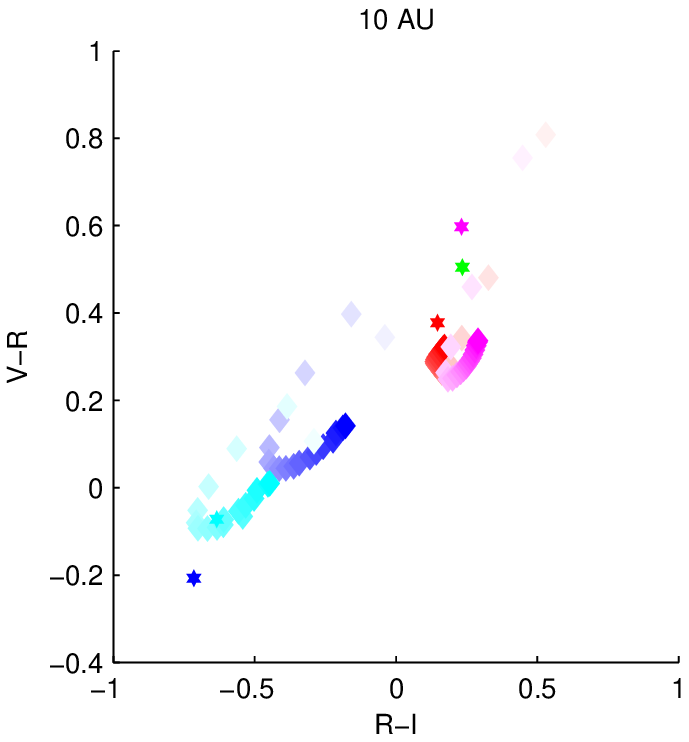}{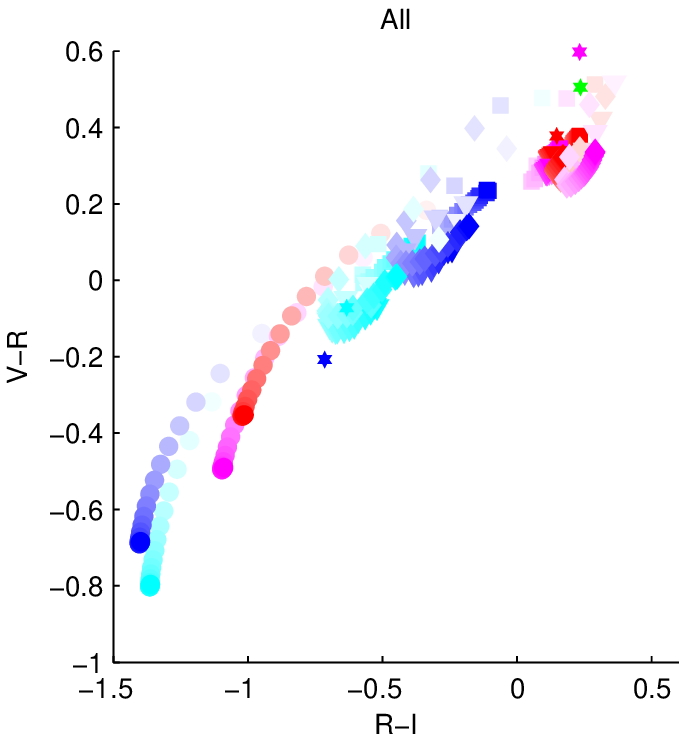}\\
\caption{Color-color comparisons. Upper left, solar system outer planets colors, calculated from Karkoschka (1994). Red = $1\times$ and magenta = $3\times$ Jupiters. Blue = $10\times$ and cyan = $30\times$ Neptunes. Upper right, 0.8 AU vs. $\alpha$, fading from $\alpha=0^{\circ}$ to $\alpha = 180^{\circ}$ in $10^{\circ}$ increments. Middle left 2 AU, middle right 5 AU, lower left 10 AU, and lower right, all. \label{vrricolor}}
\end{figure}

As mentioned in \S \ref{subs4_3}, to generate illustrative color-color diagrams of exoplanets, we use the Johnson-Morgan/Cousins UBVRI filter passbands \citep{fug95} as shown at the top of Fig.~\ref{ubvrifig}. In Fig.~\ref{vrricolor}, we first use the high-resolution solar system outer planet data from \citet{kar94} to compute the colors for Jupiter, Neptune, Saturn, Uranus, and the moon Titan (see \S \ref{subs_a9} for methods). These serve as a useful reference for comparison with the model Jupiters and Neptunes. We then compare these with the $(V-R)$ and $(R-I)$ colors of the exoplanet models. Different symbols represent different planet-star separations: circle for 0.8 AU, square for 2 AU, triangle for 5 AU and diamond for 10 AU. Red represents the $1\times$ Jupiters and magenta the $3\times$ Jupiters. Blue represents the $10\times$ Neptunes and cyan the $30\times$ Neptunes. The symbols are strongest at $\alpha = 0^{\circ}$ and fade with $\alpha$ increasing to $180^{\circ}$ in $10^{\circ}$ increments.

The cloud-free $1\times$ model at 0.8 AU exhibits behavior that is generally consistent with that in \citet{sud05}, ranging from $(R-I)$ of about $-1$ to $-0.5$ and $(V-R)$ of about $-0.4$ to 0.2 with increasing phase. The 0.8 AU $3\times$ Jupiter model and both Neptune models are also consistent with this behavior. The $1\times$ and $3\times$ Jupiter models at 2 AU and 5 AU at full phase are quite close to the real Jupiter. The $10\times$ and $30\times$ Neptune models also approach the observed Neptune, which is encouraging given that the brightness for the models are a bit darker than the observational data. We also note that for the cloudy Jupiter models, it is difficult to differentiate between the models at planet-star separations wider than 2 AU on the $(V-R)$, $(R-I)$ color-color diagrams.

\begin{figure}
\epsscale{0.8}
\plottwo{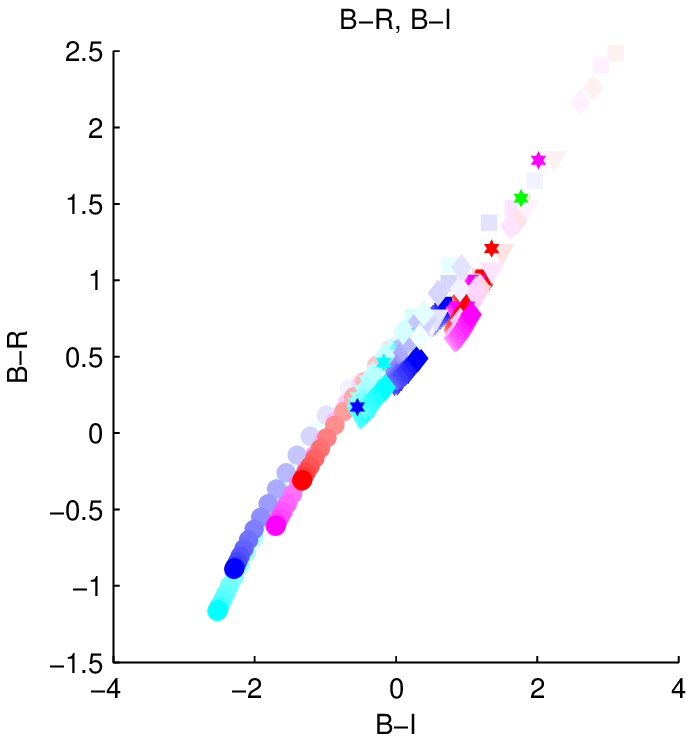}{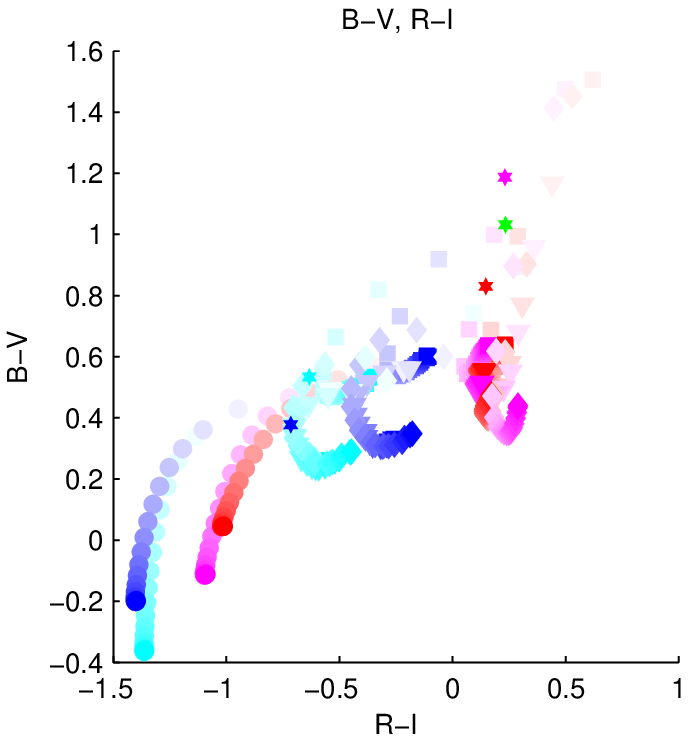}\\
\plottwo{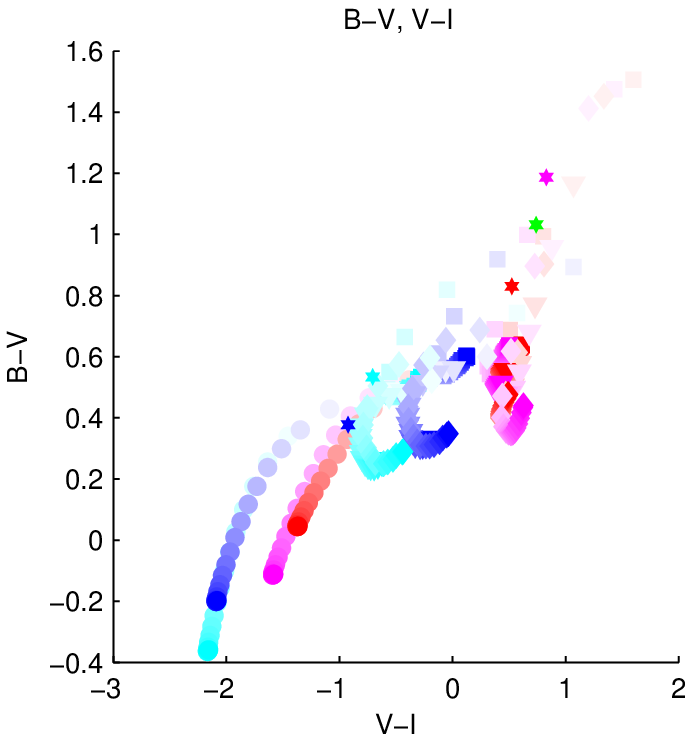}{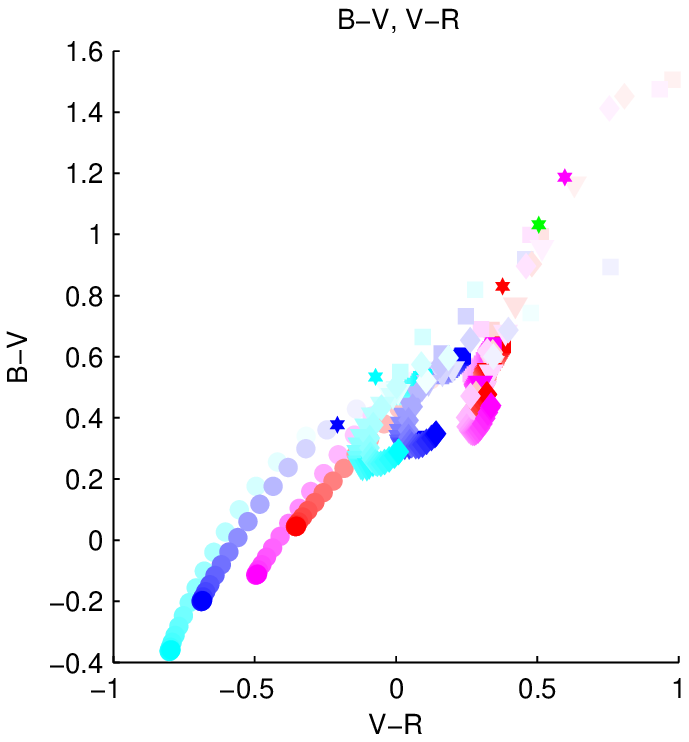}\\
\caption{Color-color comparisons, with `all' shown as in Fig.~\ref{vrricolor} lower right, but for comparisons between different color filters. The B-V, R-I comparison shows substantial horizontal and vertical offsets among the different separations and compositions.\label{othercolorcolor}}
\end{figure}

While additional work remains to be done on both generating and interpreting exoplanet albedo spectra models, another relevant question for color-color comparisons is which colors (parts of the spectrum) are most informative. Making color-color comparisons using the filters that best allow differentiation between planet models also requires further optimization. In Fig.~\ref{othercolorcolor} we show four additional color-color diagrams using different filter pairings. From this figure, it appears that the $(B-V)$, $(R-I)$ and $(B-V)$, $(V-I)$ diagrams provide the greatest spread between the planet types. These diagrams are also useful for determining how well the models capture the integrated effect of different broad regions of the spectrum in comparison to the solar system outer planet data.

It has been suggested by \citet{tra03} based on the upper left panel of Figure \ref{vrricolor} that color alone will be a definitive test of planet characteristics, for example by distinguishing terrestrial, ice-giant, and gas-giant planets.  Figures \ref{vrricolor} and \ref{othercolorcolor} suggest that such an approach is not as straightforward as it might appear.  Because planetary phase functions vary with wavelength the planet colors also depend upon phase.  Furthermore, colors vary as cloud heights vary with atmospheric temperature and composition.  Thus for the example at 5 AU in the $(V-R)$ vs. $(R-I)$ color space, the distinction between the Neptune and Jupiter-like models is slight, particularly if the phase is unknown.  We conclude that a single color-color diagram is likely inadequate to sort planets with certainty and the more photometric bandpasses that are available, the better.  Of course, the best characterization of an exoplanet will be with spectral information, preferably with $R>15$ (Figure \ref{lowresmodspec}).


\section{Conclusions and Future Work} \label{sec6}

%
\subsection{Conclusions} \label{subs6_1}


In this work, we have investigated the roles of planet-star separation, heavy element abundance, and planet phase on the reflected-light albedo spectra of Jupiter and Neptune analog exoplanets. Albedo spectra of gas giant exoplanets are sensitive to both the presence of clouds as well as to the changes in optical depth that occur as a function of planet phase, when different volumes of atmosphere are sampled corresponding to the source-observer geometry. The planet-star separation (temperature) plays a primary role in determining the cloud structure in our model exoplanet atmospheres. The abundance of heavy elements (metallicity) plays a supporting role, with clouds typically extending deeper to higher pressures and higher opacities with increasing metallicity.

\citet{mar99a}, \citet{sud00}, and \citet{sud05} have previously explored the effect of changing separation on albedo spectra of $1\,\rm M_J$ and more massive planets. We expand upon this work, developing a different model implementation to include observational geometry, exploring the effects of changing atmospheric composition and adding Neptune-mass planets.  In total, we present sixteen different model cases that span a range of parameter space consistent with the expected capabilities of early direct imaging observations of exoplanets in reflected light. The sixteen cases include a range of planet sizes (Jupiter and Neptune analogs), planet-star separations (0.8 to 10 AU), compositions (heavy element abundances between $1\times$ and $30\times$ solar), and planet phases (from $0^{\circ}$ to $180^{\circ}$ in $10^{\circ}$ increments). The exoplanet models capture cloud-free, H$_2$O, and H$_2$O + NH$_3$ cloud cases.

As previously established, clouds play a key role in controlling the brightness of a planet seen in scattered light.  Water clouds first form high in the atmosphere for planets found beyond about 1AU. With increasing planet-star separation, the water clouds are found progressively deeper in the planetary atmosphere until ammonia clouds form above them at planet-star separations $\sim5$ AU.  For cloudless planets, gaseous alkali elements, particularly Na and K, are important opacity sources along with water. Once clouds form, the main absorbing species is methane. However, the reflected light spectrum of even Jupiter itself is not purely a methane spectrum, for example, it also contains ammonia absorption features \citep{kar94, kar98}. Our computed energy budgets for our model mature gas and ice giant planets also imply that there should be substantial equator to pole temperature gradients and thus atmospheric dynamics that will differ from solar system giants.

We find that, as in the solar system, the reflected planetary spectra are sensitive to planetary phase as planets are not simply Lambertian scatterers at all wavelengths.  As a result, knowledge of the phase at which the image of an exoplanet was obtained is a key ingredient to the interpretation of the observation. Planet phase will be trivially known for planets that have also been detected by radial velocity observations.  It will be more difficult to interpret observations of exoplanets seen in only one or two images (perhaps because the planet has passed inside or outside of the inner or outer working angles of the coronagraph) and for which no RV data is available.

Ratios of albedo spectra at taken different planet phases clearly have wavelength-dependent features. Making multiple measurements of exoplanets at different phase angles, taking the ratio, and examining the relative changes in the depth of absorption features can help to characterize the vertical distribution or scattering dependences of atmospheric constituents in an exoplanet's atmosphere. The spectra generated for these exoplanet models can be used to estimate the sensitivity required to detect this type of phase-dependent effect in direct imaging observations.

As has been noted elsewhere \citep[e.g.,][]{mar97, mar07}, photochemical hazes will add a challenging aspect to any effort to interpret cool giant exoplanet spectra.  As the comparison of Jupiter and the model Jupiter in Figure \ref{karcomp} attests, the neglect of photochemical hazes is a first order discrepancy in the model.  A Neptune at 1 AU around a solar type star, with a rich brew of not only methane, but also water and $\rm H_2S$ in its atmosphere will unquestionably exhibit a complex photochemistry and thermochemistry. \citet{zah10} has only begun to explore the chemical processes that might take place in such atmospheres, but regardless we can expect that actual spectra may deviate strongly from those shown here.  If so this will add new layers of insight to that gained from the study of the scattering spectra of giant exoplanets.

Coarse spectra and color-color diagrams are a first step in characterizing an exoplanet's atmosphere. We use standard UBVRI filters to calculate the colors for each of our exoplanet models at different phase angles as well as for solar system planets Jupiter, Saturn, Saturn's satellite Titan, Neptune, and Uranus. Exoplanet colors are highly dependent on planet-star separation (clouds) and planet phase, and less sensitive to metallicity, although it may be possible to differentiate between high and low metallicity for certain filter pairings. It is important to make multiple observations and to obtain as much information as possible about the geometry of the system in order to interpret the observed spectra.

%
\subsection{Future Work} \label{subs6_2}

There are several areas that could be further developed based on this initial work. The models and methods presented here can be used to study systems using different stellar spectra and with different instrument-specific filters. Signal to noise ratio estimates can then be computed for spectral features of interest, such as Na, K, CH$_4$, NH$_3$, and H$_2$O at appropriate resolutions to help inform instrument specifications and observing plans. The exoplanet albedo spectra can be combined with model spectra of exozodiacal dust and used as input to instrument performance simulations to investigate the detectability of exoplanets as a function of wavelength in systems with substantial circumstellar debris structure.

Additional exoplanet model `points' in the metallicity and planet-star separation space can be considered. Given the relatively small effect of increasing from $10\times$ to $30\times$ solar metallicity on the albedo spectra for Neptunes at separations larger than 2 AU, and yet the noticeable effect of increasing from $1\times$ to $3\times$ solar metallicity for the Jupiters, it would be interesting to adjust abundances and planet-star separations to determine where the effect begins to diminish.

Cloud formation has a strong effect on albedo spectra. It would be instructive to consider the effect of different sedimentation ratios for these exoplanet models. It would also be interesting to adjust the vertical extent of the clouds to investigate the effect of longer path lengths through the clouds at different phase angles and for different vertical distributions. A similar investigation to compare the absorption features of well-mixed versus stratified or pressure-dependent atmospheric species as a function of phase would also be useful in preparing to characterize exoplanets with direct imaging observations. These types of investigations would benefit from updating the current cloud model to output the full anisotropic phase function in addition to the asymmetry parameter and single scattering albedo.

In this work, we restricted the system geometry to the simple case of circular orbits with no other variation in the Keplerian elements (such as eccentricity, inclination, or obliquity). Investigation of more complex orbits and the effect of the changing planet-star separation through the orbit on the T(P) structure of the atmosphere would be informative. The time-dependence and evolution of an exoplanet's atmosphere through a more complex orbit could have a substantial effect on the observed albedo spectra. The obliquity of a planet and the potential for seasonal change would further complicate the interpretation of albedo spectra. 

With the ability to create many facets over an illuminated sphere, we could use models for several types of facets to paint a portrait of a non-uniform exoplanet on the sphere. The contributions from the different facets would then be integrated to generate the albedo spectra. For example, we could create facets that correspond to banded structure, or other variations in cloud topology.

\appendix


\section{Appendix A} \label{App}

\subsection{Color magnitudes and Filters} \label{subs_a9}

Relative color-color comparisons, such as $u-v$, can be made using the filter response, the albedo spectrum, and parent star spectrum:

\begin{equation}
u-v = -2.5 \log_{10}\left(\frac{\int_{\lambda}\rm R_{\rm U, \lambda}\rm A_{\alpha, \lambda} F_{\odot, \lambda}d\lambda}{\int_{\lambda}\rm R_{\rm V, \lambda}\rm A_{\alpha, \lambda} F_{\odot, \lambda}d\lambda}\right)
\label{eq_a91}
\end{equation}

\noindent where $\rm R_{\rm U, \lambda}$ and $\rm R_{\rm V,\lambda}$ are the U and V filter responses as shown in Fig.~\ref{ubvrifig}, $F_{\odot,\lambda}$ is the spectrum of the parent star, and $\rm A_{\alpha,\lambda}$ is the albedo spectrum of the exoplanet at some phase $\alpha$. Usually it is assumed that $\alpha = 0^{\circ}$ and the geometric albedo spectrum is used, but in this work we are interested in changes with $\alpha$ and use the albedo spectra corresponding to different phases.

Planets in our solar system have apparent magnitudes that assume the planets are illuminated by the sun and observed from Earth. For the purpose of comparison, we can take the approach of calculating magnitudes for the exoplanet models as though they were also in the solar system, using their geometric albedo spectrum, the solar spectrum, the filter response, and the reference spectrum. For example, to calculate $V$ for an exoplanet,

\begin{equation}
V_{\rm exoplanet} = -2.5 \log_{10}\left(\frac{\int_{\lambda }\rm R_{\rm V, \lambda}\rm A_{\alpha, \lambda} F_{\odot, \lambda}d\lambda}{\int_{\lambda}\rm R_{\rm V, \lambda} F_{\rm ref, \lambda}d\lambda}\right) + 5\log_{10}\left(\frac{\Delta \cdot r}{{\rm{R_p}} r_0} \right)
\label{eq_a92}
\end{equation}

\noindent where $F_{\rm ref,\lambda}$ is the reference spectrum, $\Delta$ is the distance from the observer on Earth to the planet, $r$ is the distance from the parent star to the planet (both in AU), $\rm R_p$ is the radius of the planet (in AU) and $r_0$ is 1 AU.

To go from planetary to absolute magnitudes (at 10 pc) requires adding $5\log_{10}\left(10\cdot 206,265\right)$, where 1 pc $\sim206,265$ AU.

\section{Acknowledgments}

The authors thank the referee for useful comments and suggestions that improved the manuscript. We thank Richard Freedman for providing the opacity tables used in this work, and Katharina Lodders for the elemental abundances and the condensation curves. We thank Olivier Guyon for making the initial version of the PIAA coronagraph simulation available for our use. We thank Chris McKay for several useful discussions regarding the development of the radiative transfer model. K. C. is supported by an appointment to the NASA Postdoctoral Program at NASA Ames, administered by Oak Ridge Associated Universities through a contract with NASA. M. M. and J. F. acknowledge support from the NASA Planetary Atmospheres program.



\clearpage



\clearpage

\clearpage

\clearpage

\clearpage

\clearpage

\clearpage

\clearpage

\clearpage

\clearpage

\clearpage

\clearpage

\clearpage

\clearpage

\clearpage

\clearpage

\begin{figure}
\epsscale{0.6}
\plotone{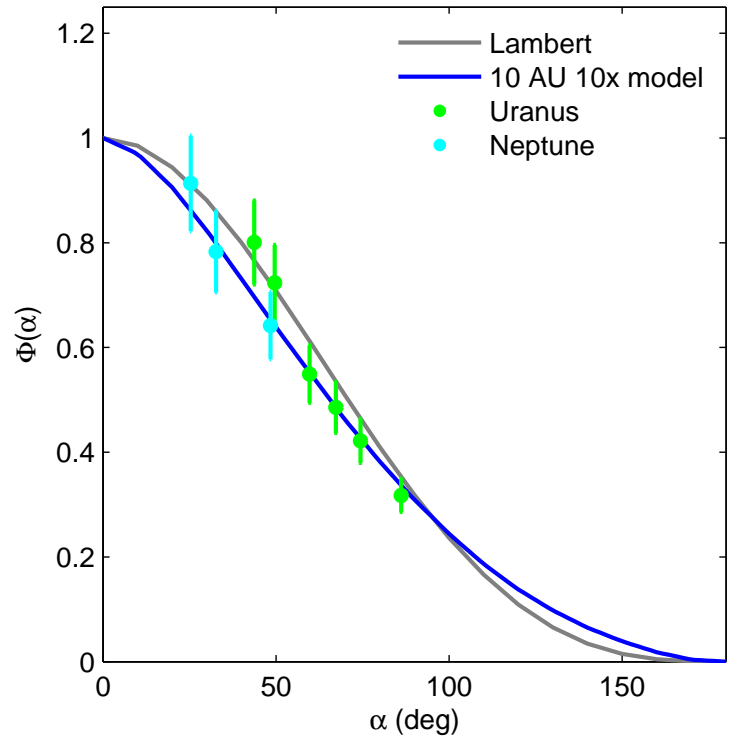}
\caption{Comparison of our Neptune 10 AU at 10$\times$ solar metallicity model with data at higher phase angles from Voyager 1 observations of Uranus and Neptune \citep{pol86}. \label{pollackfig}}
\end{figure}

\clearpage

\clearpage

\clearpage

\clearpage

\clearpage

\clearpage

\begin{table}
\begin{center}
\caption{Description of the model gas and ice giants used as inputs to the albedo spectra simulation. The separations and cloud types are noted. $T_{\rm{int}}$ is a parameterizaton of the flux from the interior of the planet, due to its cooling with time (temperature in the absence of any incident flux).  $f_{sed}$ is the ratio of the microphysical sedimentation flux to the eddy sedimentation flux. \label{tab1}}

\begin{tabular}{c | c c c | c c c | c  c }
 & \multicolumn{3}{| c |}{Jupiters} & \multicolumn{3}{| c |}{Neptunes}  & &  \\
& \multicolumn{3}{| c |}{$g = 25$ ms$^{-2}$} & \multicolumn{3}{| c |}{$g = 10$ ms$^{-2}$} & & \\
Separation & $T_{\rm int}$ (K) & \multicolumn{2}{c |}{$T_{\rm eff}$ (K)} & $T_{\rm int}$ (K) & \multicolumn{2}{c |}{$T_{\rm eff}$ (K)} & Clouds & $f_{\rm sed}$\\
& & $1\times$ & $3\times$ & & $10\times$ & $30\times$ & & \\
\tableline
0.8 AU & 100 & 274 & 282 &  50 & 285 & 292  & None & - \\
2 AU & 100 & 123 & 126 & 50 & 134 & 148 & H$_2$O & 6 \\
5 AU & 100 & 115 & 119 & 50 & 110 & 112 & H$_2$O, NH$_3$ & 10 \\
10 AU & 100 & 107 & 109 & 50 & 81 & 82 & H$_2$O, NH$_3$ & 10 \\
\end{tabular}
\end{center}
\end{table}

\clearpage
\begin{table}
\begin{center}
\caption{Characteristic `effective radii' ($\mu$m) of cloud condensate particles predicted by the \citet{ack01} model. The parameter `effective radius' is used for comparative purposes to summarize the level-dependent distribution of radii actually used in the model\label{tab1b}}
\begin{tabular}{c | c c | c c | c c | c c |}
 & \multicolumn{4}{| c |}{Jupiters} & \multicolumn{4}{| c |}{Neptunes}  \\
& \multicolumn{2}{| c |}{1 $\times$} & \multicolumn{2}{| c |}{3 $\times$} & \multicolumn{2}{| c |}{1 $\times$} & \multicolumn{2}{| c |}{3 $\times$} \\
Separation & H$_2$O & NH$_3$& H$_2$O & NH$_3$ & H$_2$O & NH$_3$& H$_2$O & NH$_3$\\
\tableline
0.8 AU &  & - & - & - & - & - & - & - \\
2 AU & 69 & - & 72 & - & 52 & - & 53 & - \\
5 AU & 102 & 56 & 103 & 61 & 72 & 61 & 73 & 67 \\
10 AU & 100 & 74 & 102 & 87 & 80 & 71 & 80 & 73 \\
\end{tabular}
\end{center}
\end{table}

\clearpage
\begin{table}
\begin{center}
\caption{Approximate wavelengths of optical absorption features noted in our gas giant exoplanet models. The references in this table refer to work by other researchers who have noted these features. The use of `weak' refers only to the appearance of the spectral features in this work; for example, resolution here is not sufficient to resolve expected doublet absorption features.\label{tab2}}
\begin{tabular}{c c c}
Approximate $\lambda$ ($\mu$m) & Species & Reference \\
\tableline
0.40 & K & 3, 7 \\		
0.46 & CH$_4$ &1 \\	
0.48 & CH$_4$ & 1 \\	
0.54 & CH$_4$ &1 \\	
0.59	& Na `doublet' & 3, 4, 5, 7 \\		
0.62	& CH$_4$ & 1 \\		
0.65	& H$_2$O weak & 3, 6 \\		
0.73	& H$_2$O weak & 3, 6 \\		
0.77	& K `doublet' weak & 3, 4, 5, 7\\		
0.78	& CH$_4$ & 1 \\
0.79	& CH$_4$ & 1 \\		
0.83	& H$_2$O weak & 2, 3, 6 \\		
0.84	& CH$_4$ & 1 \\		
0.86	& CH$_4$ & 1 \\		
0.89	& CH$_4$ & 1 \\		
0.91	& CH$_4$ & 1 \\
0.94	& H$_2$O & 1, 2, 3, 4, 6 \\
0.99	& CH$_4$ & 1 
\end{tabular}
\tablenotetext{1}{\citet{kar94}} 
\tablenotetext{2}{\citet{mar99a}}
\tablenotetext{3}{\citet{sud00}}
\tablenotetext{4}{\citet{bur04}}
\tablenotetext{5}{\citet{for08b}}
\tablenotetext{6}{R. Freedman, \textit{pers. comm.}}
\tablenotetext{7}{NIST atomic spectra database, http://www.nist.gov/physlab/data/asd.cfm}
\end{center}
\end{table}

\clearpage
\begin{table}
\begin{center}
\caption{Geometric albedos ($\rm A_g$) of the exoplanet model Jupiter and Neptunes in this work, for each of the B, V, R, and I filters shown in Fig.~\ref{ubvrifig} and the albedo spectra model described in \S \ref{subs3_3} and Appendix \ref{App}. \label{tab3}}
\begin{tabular}{l  c  c  c  c }
& B & V & R & I \\
Model & $ \rm A_g$  & $ \rm A_g$  & $ \rm A_g$ & $\rm A_g$  \\
\tableline
0.8 AU, Jupiter $1 \times$ & 0.556 & 	0.322 & 	0.156 & 	0.047\\
0.8 AU, Jupiter $3 \times$ &0.482 & 	0.241 & 	0.102 & 	0.029\\
0.8 AU, Neptune $10 \times$ &0.455 & 	0.209 & 	0.074 & 	0.016\\
0.8 AU, Neptune $30 \times$ &0.359 & 	0.142 & 	0.045 & 	0.010\\
\tableline
2 AU, Jupiter $1 \times$ &0.733 & 	0.742 & 	0.727 & 	0.673\\
2 AU, Jupiter $3 \times$ &0.758 & 	0.766 & 	0.726 & 	0.630\\
2 AU, Neptune $10 \times$ &0.744 & 	0.728 & 	0.616 & 	0.420\\
2 AU, Neptune $30 \times$ &0.735 & 	0.674 & 	0.498 & 	0.267\\
\tableline
5 AU, Jupiter $1 \times$ &0.609 & 	0.567 & 	0.531 & 	0.453\\
5 AU, Jupiter $3 \times$ &0.567 & 	0.506 & 	0.458 & 	0.388\\
5 AU, Neptune $10 \times$ &0.434 & 	0.326 & 	0.238 & 	0.145\\
5 AU, Neptune $30 \times$ &0.430 & 	0.303 & 	0.191 & 	0.090\\
\tableline
10 AU, Jupiter $1 \times$ &0.450 & 	0.386 & 	0.358 & 	0.313\\
10 AU, Jupiter $3 \times$ &0.316	 & 0.260 & 	0.246 & 	0.240\\
10 AU, Neptune $10 \times$ &0.388 & 	0.295 & 	0.228 & 	0.147\\
10 AU, Neptune $30 \times$ &0.388 & 	0.279 & 	0.189 & 	0.096\\
\end{tabular}
\end{center}
\end{table}

\clearpage
\begin{table}
\begin{center}
\caption{Phase integrals ($q$) of the exoplanet model Jupiter and Neptunes in this work, using the filters shown in Fig.~\ref{ubvrifig} and the albedo spectra model described in \S \ref{subs3_3} and Appendix \ref{App}. \label{tab4}}
\begin{tabular}{l  c  c  c  c }
Model & $q_{\rm B}$  & $q_{\rm V}$  & $q_{\rm R}$ & $q_{\rm I}$  \\
\tableline
0.8 AU, Jupiter $1 \times$ &1.470 &	1.582 &	1.693 &	1.764 \\
0.8 AU, Jupiter $3 \times$ &1.501 &	1.636 &	1.739 &	1.784 \\
0.8 AU, Neptune $10 \times$ &1.509 &	1.639 &	1.725 &	1.748 \\
0.8 AU, Neptune $30 \times$ &1.561 &	1.698 &	1.762 &	1.772 \\
\tableline
2 AU, Jupiter $1 \times$ &1.367 &	1.329 &	1.305 &	1.281 \\
2 AU, Jupiter $3 \times$ &1.366 &	1.328 &	1.300 &	1.267 \\
2 AU, Neptune $10 \times$ &1.397 &	1.356 &	1.313 &	1.251 \\
2 AU, Neptune $30 \times$ &1.400 &	1.357 &	1.307 &	1.243 \\
\tableline
5 AU, Jupiter $1 \times$ &1.316 &	1.286 &	1.271 &	1.261 \\
5 AU, Jupiter $3 \times$ &1.297 &	1.270 &	1.257 &	1.248 \\
5 AU, Neptune $10 \times$ &1.523	 &1.529 &	1.463 &	1.335 \\
5 AU, Neptune $30 \times$ &1.523	 &1.530 &	1.462 &	1.349 \\
\tableline
10 AU, Jupiter $1 \times$ &1.372 &	1.312 &	1.277 &	1.253 \\
10 AU, Jupiter $3 \times$ &1.407 &	1.327 &	1.275 &	1.240 \\
10 AU, Neptune $10 \times$ &1.544 &	1.527 &	1.436 &	1.275 \\
10 AU, Neptune $30 \times$ &1.543 &	1.527 &	1.433 &	1.279 \\
\end{tabular}
\end{center}
\end{table}


\clearpage
\begin{table}
\begin{center}
\caption{Calculated Bond albedos ($\rm A_B$) of the exoplanet model Jupiter and Neptunes in this work, integrating from 0.35 to 2.5 $\mu$m over geometric albedo spectra and phase integrals from the albedo spectral model described in \S \ref{subs3_3} and Appendix \ref{App}. Contributions from longer wavelengths than 2.5 $\mu$m to the Bond albedo are smaller than 5\%. \label{tab6}}
\begin{tabular}{l  c  c  c  c }
Model & $\rm A_B$  \\
\tableline
0.8 AU, Jupiter $1 \times$ & 0.408 \\
0.8 AU, Jupiter $3 \times$ & 0.333 \\
0.8 AU, Neptune $10 \times$ & 0.297 \\
0.8 AU, Neptune $30 \times$ & 0.229 \\
\tableline
2 AU, Jupiter $1 \times$ & 0.914 \\
2 AU, Jupiter $3 \times$ & 0.899 \\
2 AU, Neptune $10 \times$ & 0.787 \\
2 AU, Neptune $30 \times$ & 0.685 \\
\tableline
5 AU, Jupiter $1 \times$ & 0.675 \\
5 AU, Jupiter $3 \times$ & 0.592 \\
5 AU, Neptune $10 \times$ & 0.407 \\
5 AU, Neptune $30 \times$ & 0.366 \\
\tableline
10 AU, Jupiter $1 \times$ & 0.489 \\
10 AU, Jupiter $3 \times$ & 0.348 \\
10 AU, Neptune $10 \times$ & 0.378 \\
10 AU, Neptune $30 \times$ & 0.344 \\
\end{tabular}
\end{center}
\end{table}

\end{document}